
*****  ATTACHMENT: art5s.tex *****


\font \titlefont   = cmbx10  scaled 1578
\font \pargfont    = cmbx10  scaled 1316
\font \abstractfont= cmbx10  scaled \magstephalf
\font \subpargfont = cmbx10  scaled \magstephalf
\font \authorfont  = cmr10   scaled \magstephalf
\font \addressfont = cmr10   scaled \magstephalf
\font \reffont     = cmbx10
\font \lemmafont   = cmbx10
\font \propfont    = cmbx10
\font \prooffont   = cmsl10
\font \remarkfont  = cmsl10
\font \examplefont = cmsl10


\magnification = \magstephalf
\raggedbottom


\newcount \formulanr   \formulanr = 0
\newcount \pargnr      \pargnr    = 0
\newcount \subpargnr   \subpargnr = 0
\newcount \lemmanr     \lemmanr   = 0
\newcount \propnr      \propnr    = 0
\newcount \examplenr   \examplenr = 0
\newcount \theoremnr   \theoremnr = 0
\newcount \ampnr       \ampnr     = 0
\newcount \figurenr    \figurenr  = 0


\def \referenties {\bigskip \centerline {\reffont References}
\medskip\immediate\write16{References}}
\def \refbook#1#2#3#4{\item{\lbrack #1\rbrack } #2:\ {#3}.\ #4}

\def \refart#1#2#3#4#5#6{\item{\lbrack #1\rbrack } #2:\ {#3}.\ #4\ ${\bf
{#5}}$,\ #6}
\def \refpre#1#2#3{\item{\lbrack #1\rbrack } #2:\ {\it #3}. (Preprint)}

\def \referentie#1{{\rm [#1]}}


\def \Abstract {\bigskip{\abstractfont Abstract}\smallskip\par}
\def \ackn#1 {\medskip\noindent{\remarkfont Acknowledgments.}\ #1}
\def \address #1 {\centerline {\addressfont #1} \vskip 2truecm}
\def \Amplification{\advance\ampnr by 1
                    \noindent{\remarkfont Amplification \the \ampnr}\par}
\def \authors #1 {\centerline {\authorfont #1} \bigskip}
\def \contents#1{\bigskip\bigskip{\noindent\pargfont #1}\bigskip}
\def \Corollary #1 {\bigskip\noindent\advance\theoremnr by 1
                   {\propfont Corollary \the\theoremnr\quad }
                   {\it #1}\medskip}
\def \endparg{\eject}

\def \endproof {{\nobreak\hfill {\vrule height4pt width3pt
depth2pt}}\medskip\par}

\def \endtitle {\vskip 1.5 truecm}
\def \Example{\bigskip\advance\examplenr by 1\noindent{\examplefont Example
}\par}
\def \formula {\global \advance \formulanr by 1 \eqno (\the \formulanr)}
\def \introtitle#1{\bigskip\noindent{\it #1}\nobreak\medskip\nobreak}
\def \Lemma #1 {\bigskip\noindent\advance\theoremnr by 1
                {\lemmafont Lemma \the\theoremnr.\quad }
                {\it #1}\medskip}
\def \parg #1 {\eject\noindent\advance\pargnr by 1\subpargnr=0
              {\pargfont  \the\pargnr. #1} \bigskip
               \immediate\write16{Paragraph \the\pargnr}}
\def \Proof{\noindent{\prooffont Proof}\nobreak\par\nobreak}
\def \Proofof#1{\noindent{\prooffont Proof of #1}\nobreak\par}
\def \received #1 {\centerline {\addressfont #1} \bigskip}

\def \subparg #1 {\bigskip\noindent\advance\subpargnr by 1
                 {\subpargfont \the\pargnr.\the\subpargnr. #1}
                 \nobreak\smallskip\nobreak
                 \message{(Par. \the\pargnr.\the\subpargnr)}}
\def \title #1 {\centerline {\titlefont #1}}

\long\def \Proposition #1 {\bigskip\noindent\advance\theoremnr by 1
                {\propfont Proposition \the\theoremnr\quad }
                {\it #1}\medskip}
\long \def \Theorem #1 {\bigskip\noindent\advance\theoremnr by 1
                {\propfont Theorem \the \theoremnr\quad }
                {\it #1}\medskip}


\def \A{{\cal A}}
\def \AFd{{\cal A}_{F,d}}
\def \AFdC{{\cal A}_{F,d}^{\rm\Complex}}
\def \DFd{{\cal D}_{F,d}}
\def \EFd{{\cal E}_{F,d}}
\def \Fu{F_{(u)}}
\def \G#1{\Gamma_{#1}}
\def \GB{\bar\Gamma}
\def \GBF{\bar\Gamma_F}
\def \GF{\Gamma_F}
\def \l {\lambda}
\def \m {\mu}
\def \min#1{\left\lbrack#1\right\rbrack_-}
\def \otheta{{\cal O}}

\def \Poisson#1{\left\{#1\right\}_d^\varphi}
\def \poisson#1{\left\{#1\right\}_d}
\def \Poissonnil{\Poisson{\cdot\,,\cdot}}
\def \poissonnil{\poisson{\cdot\,,\cdot}}
\def \plus#1{\left\lbrack#1\right\rbrack_+}
\def \Rd {\R{2d}}
\def \sub{{1\leq i,j\leq d}}


\def \Id{{\rm Id}}
\def \Jac{{\rm Jac}}

\def \mod{\ {\rm mod}\,}

\def \S{{\cal S}}
\def \Sym#1{\hbox{\rm Sym}^{#1}}
\def \Symd{\Sym d}


\font\cmss=cmss10 \font\cmsss=cmss10 at 7pt
\def \inbar{\,\vrule height1.5ex width.4pt depth0pt}

\def \Complex{{\relax\hbox{$\inbar\kern-.3em{\rm C}$}}}
\def \C#1{\Complex^{#1}}
\def \PC{\relax{\rm I\kern-.18em P}}
\def \CP#1{\PC^{#1}}
\def \Z{\relax\ifmmode\mathchoice
       {\hbox{\cmss Z\kern-.4em Z}}{\hbox{\cmss Z\kern-.4em Z}}
       {\lower.9pt\hbox{\cmsss Z\kern-.4em Z}}
       {\lower1.2pt\hbox{\cmsss Z\kern-.4em Z}}\else{\cmss Z\kern-.4em Z}\fi}
\def \Real{\relax{\rm I\kern-.18em R}}
\def \R#1{\Real^{#1}}
\def \N{\relax{\rm I\kern-.18em N}}
\def \F{\relax{\rm I\kern-.18em F}}


\newdimen\unit
\def\putatpos#1 #2 #3{\vbox to0pt{\kern-#2\unit
                      \hbox{\kern#1\unit#3}\vss}\nointerlineskip}
\unit=\baselineskip


\def \nameformula#1{}
\def \formulaname#1{}
\def \namesection#1{}
\def \sectionname#1{}
\def \figurename#1{}
\def \namefigure#1{}
\def \nameproposition#1{}
\def \propositionname#1{}
\def \namelemma#1{}
\def \lemmaname#1{}

%
%

\nopagenumbers
\hrule height0pt
\vskip 4 true cm
\title {Integrable systems and symmetric products of curves}
\endtitle
\authors {Pol Vanhaecke}
\Abstract
We show how there is associated to each non-constant polynomial $F(x,y)$ a
completely integrable system with polynomial invariants on $\Rd$ and on
$\C{2d}$ for each
$d\geq1$; in fact the invariants are not only in involution for one
Poisson bracket, but for a large class of polynomial
Poisson brackets, indexed by the
family of polynomials in two variables.
We show that the complex invariant
manifolds are isomorphic to affine parts
of $d$-fold
symmetric products of a deformation of the algebraic curve $F(x,y)=0$,
and derive the structure of the real invariant manifolds from it.
We also
exhibit Lax equations for the hyperelliptic case (i.e., when $F(x,y)$ is of
the form $y^2+f(x)$) and we show that in this case the invariant manifolds are
affine parts of distinguished (non-linear) subvarieties of the Jacobians of
the curves. As an application the geometry of the H\'enon-Heiles hierarchy ---
a family of superimposable integrable polynomial potentials
on the plane --- is revealed
and Lax equations for the hierarchy are given.
\vskip 4 true cm
\itemitem{Address:} Universit\'e des Sciences et Technologies de Lille
\itemitem{}         U.F.R. de Math\'ematiques
\itemitem{}         59655 Villeneuve D'Ascq Cedex
\itemitem{}         France
\itemitem{E-mail:}  Vanhaeck@gat.Univ-Lille1.fr
\eject
\contents{Table of contents}
\item{1} Introduction
\bigskip
\item{2}{The systems and their integrability}
\medskip
\item{2.1}{Notation}
\smallskip
\item{2.2}{The compatible Poisson structures $\Poissonnil$}
\smallskip
\item{2.3}{Polynomials in involution for $\Poissonnil$}
\bigskip
\item{3}{The geometry of the invariant manifolds}
\medskip
\item{3.1}{The invariant manifolds $\AFd$ and {\rm $\AFdC$}}
\smallskip
\item{3.2}{The structure of the complex invariant manifolds {\rm $\AFdC$}}
\smallskip
\item{3.3}{The structure of the real invariant manifolds $\AFd$}
\smallskip
\item{3.4}{Compactification of the complex invariant manifolds {\rm $\AFdC$}}
\bigskip
\item{4}{The hyperelliptic case}
\medskip
\item{4.1}{Lax equations}
\smallskip
\item{4.2}{$\AFdC$ as strata of hyperelliptic Jacobians}
\smallskip
\item{4.3}{The H\'enon-Heiles hierarchy}
\eject

\footline {\hss\tenrm\folio\hss}
\pageno = 1

\parg{Introduction}
Finite-dimensional integrable systems first appeared in the works of
Euler (1758), Lagrange (1766), Jacobi (1836), Liouville (1846) and
Kowalewski (1889).
They were given as systems of (non-linear) differential equations
describing the motion of a mechanical system, having a sufficient number
of integrals. Their investigation was based on the fact that the
equations and the integrals were polynomials (in some coordinates) and
led, in all cases considered, to an explicit integration of these
equations in terms of (hyperelliptic) theta functions, well-known in
algebraic geometry. Their work clearly showed the rich interplay between
the theory of Riemann surfaces/algebraic curves (which was in that time
thought of as a chapter in complex analysis) and mechanics.
During the first half of the present century however,
algebraic geometry was refounded
and became ever more abstract, while in the theory of mechanical
systems, generic smooth dynamical systems were gaining interest
(as opposed to integrable ones). So both theories got separated, and
integrable systems --- which were at the core of this intimate
relationship --- faded away from the picture.
\medskip
The interest in both integrable systems and their connection to
algebraic geometry revived in the early seventies; many integrable
systems were found as finite-dimensional solutions of certain
(integrable) partial differential equations (such as the well-known
Korteweg-de Vries equation) and they were again integrated in terms of
theta functions. Their study lead in particular to the concept of an
algebraic completely integrable system (a.c.i.\ system). Shortly, such
a system is an integrable system which has a complexification for which
the invariant manifolds (the smooth level sets of the integrals) are
(open subsets of) complex algebraic tori, and the flow (run with complex time)
is
linear on these tori (see \referentie{AvM1}, \referentie{Mu}). Algebraic
geometry has been shown to be a useful tool for the study of a.c.i.
systems and a solution to some problems in algebraic geometry was found
by using an a.c.i.\ system.
\medskip
The present paper deals with a (new) class of integrable systems which
(apart from an exceptional case, specified below) do not fall in the
class of a.c.i.\ systems, but yet they have a natural complexification
and their geometry is most naturally described by using algebraic
geometry. Apart from their construction, their geometry will be analysed
in detail and we will show how these systems can be used to explain the
geometry of several known integrable systems which are not a.c.i.
\introtitle{1.1 Poisson structures on $\Rd$}
On $\Rd$ with coordinates $(u_1,\dots,u_d,$ $v_1,\dots,v_d)$ we show in
Section 2.2 that there
corresponds in a natural way
to any non-zero polynomial $\varphi(x,y)\in\Real[x,y]$ a
Poisson bracket $\Poissonnil$, which is given by
  $$ \eqalign{\Poisson{u({\l}),u_j}&=\Poisson{v({\l}),v_j}=0,\cr
              \Poisson{u({\l}),v_j}&=\Poisson{u_j,v({\l})}=\varphi
              (\l,v(\l))\plus{u({\l})\over {\l}^{d-j+1}}\mod u(\l),
                    \qquad 1\leq j\leq d,}
     \formula
  $$
\formulaname{Poisson_intro}where $u(\l)=\l^d+u_1\l^{d-1}+\cdots+u_d$ and
$v(\l)=v_1\l^{d-1}+\cdots+v_d$; also $\plus{R(\l)}$ denotes the
polynomial part of a rational function $R(\l)$ and $f(\l)\mod g(\l)$ is
the rest obtained when dividing $f(\l)$ by $g(\l)$. The map
$\varphi\mapsto\Poissonnil$ is clearly a linear map, which is moreover
injective, since the Poisson structures obtained are of maximal rank
except for $\phi=0$. If $\phi(x,y)$ is a constant, say $\phi(x,y)=1$,
then the bracket $\poissonnil=\{\cdot\,,\,\cdot\}_d^1$ is given by the
following matrix $P$ of Poisson brackets:
  $$ P=\pmatrix{0& U\cr -U&0}\quad \hbox{where}\quad
      U=\pmatrix{0&0&\cdots&0&1\cr
        0&0&\cdots&1&u_1\cr
        \vdots&\vdots&&\vdots&\vdots\cr
        0&1&\cdots&u_{d-3}&u_{d-2}\cr
        1&u_1&\cdots&u_{d-2}&u_{d-1}}.
  $$
Thus,
(1)\nameformula{Poisson_intro} provides us
with a large class of Poisson structures on $\Rd$, which are in fact
polynomial, i.e., all brackets of the coordinates $u_i$ and $v_j$ are
polynomials; moreover they are all compatible in a sense explained in the
text.
\introtitle{1.2 Integrable systems on $\Rd$}
What is remarkable is that these Poisson structures have a very large
class of integrable systems in common, namely one corresponding to every
polynomial $F(x,y)\in\Real[x,y]\setminus\Real[x]$. To describe these,
let $F(x,y)$ be such a polynomial  and
expand $F(\l,v(\l))\mod u(\l)$ as a polynomial in $\l$ (of degree
$d-1$):
  $$ F({\l},v({\l})) \mod u({\l})=H_1\l^{d-1}+H_2\l^{d-2}+\cdots+H_d;.
  $$
Remark that $H_1,\dots,H_d$ are polynomials in $u_i$ and $v_j$. The main
result, established in Section 2.3, is that these polynomials
Poisson commute for all brackets $\Poissonnil$ on $\Rd$, that is
  $$\Poisson{H_i,H_j}=0\quad\hbox{for all $1\leq i,j\leq d$ and
        $\varphi(x,y)\in\Real[x,y].$}
  $$
Since $H_1,\dots,H_d$ are independent, the conclusion is that for any
polynomial $F(x,y)\in\Real[x,y]\setminus\Real[x]$ we have an
integrable system on the Poisson manifold
$\left(\Rd,\Poissonnil\right)$, where $d\geq1$ is arbitrary and
$0\neq\varphi(x,y)\in\Real[x,y]$ is an arbitrary polynomial dictating
the Poisson structure, and our construction is totally explicit.
\smallskip
Since everything in our construction is polynomial, these systems have a
natural complexification as complex integrable systems on the Poisson
manifold $\left(\C{2d},\Poissonnil\right)$, where the Poisson structure
$\Poissonnil$ is now a holomorphic one.
\introtitle{1.3 The geometry of the systems}
The meaning of the polynomial $F(x,y)$ and the need for considering the
complexified system becomes apparent in Section 3, when we study
(for generic values
of $c_i$) the level sets $\AFd=\{P\in\Rd\mid H_i(P)=c_i\}$, which are
preserved by the flows of the vector fields associated to all $H_i$.
Namely we will show in Section 3.2 that the complex invariant set (lying over
$0$)
  $$\AFdC=\{(u(\l),v(\l))\in\C{2d}\mid H_{F,d}(u(\l),v(\l))=0\}
  $$
is (biholomorphic to) an affine part of the $d$-fold symmetric product
of the plane algebraic curve $\GF\subset\C2$, defined by $F(x,y)=0$
($\GF$ is supposed generic here, i.e., smooth); a similar description of
the structure of the
other complex invariant sets (lying over $(c_1,\dots,c_d)$) follows at
once. The real invariant sets being the fixed points on $\AFdC$ of the
complex conjugation map, we obtain in Section 3.3
a description of $\AFd=\AFdC\cap\Rd$ as the set of all $d$-tuples in
$\AFdC$, consisting only of real points and points which appear in
complex conjugated pairs. We will show how this leads to an
explicit description of the topology
of the invariant manifolds $\AFd$, which are in general neither tori nor
cylinders. The compactification of the complex invariant manifolds, of
major interest in several studies in this field, is discussed in Section
3.4: it turns out that in general a smooth compactification of the complex
level
manifolds, such that the vector fields of the system extend in a
holomorphic way to them, does {\sl not} exist.
\introtitle{1.4 The hyperelliptic case}
The special case where $F(x,y)$ is of the form $F(x,y)=y^2+f(x)$ will be
considered in more detail in Section 4. Then the vector fields
$X^\varphi_{H_i}$ of the integrable
system can be written as Lax equations
  $$X^\varphi_{H_i}A({\l})=\left\lbrack A({\l}),\plus{B_i({\l})}\right\rbrack,
  $$
where
  $$ A({\l})=\pmatrix{v({\l})& u({\l})\cr -\plus{F(\l,v(\l))\over u(\l)}
               &-v({\l})}\quad\hbox{and}\quad
           B_i({\l})={\varphi(\l,v(\l))
                \over u({\l})}\plus{u({\l})\over {\l}^{d-i+1}}A(\l);
     \formula
  $$
\formulaname{hyperelliptic_Lax}see Section 4.1.
The geometry of these systems can be related
to that of the Jacobian of the curve $\GF$. In particular, in the very
special case that $\varphi(x,y)=1$ and $d=\hbox{genus }(\GF)$ the
manifold $\AFdC$ is an affine part of the Jacobian of $\GF$, the flow of
the vector fields is linear and the system is algebraic completely
integrable. If $d<\hbox{genus }(\GF)$ then $\AFdC$ is interpreted as a
very special non-linear subvariety of the Jacobian of $\GF$.
\smallskip
The
geometry of several
integrable systems, such as the H\'enon-Heiles hierarchy and its
generalisations (in different aspects), the (generalised) Gaudin magnet,
the discrete self-trapping timer,$\dots$, can be described in very much
detail by using our systems. We will show in Section 4.3 quite detailed how
this is
done for the H\'enon-Heiles hierarchy, which consists of a
family of (superimposable) integrable potentials on the
plane. For the other examples one proceeds in a completely analogous
way.
\medskip
%
\endparg
\parg{The systems and their integrability}
\sectionname{systems}In
this section we describe our basic construction, which associates to a
pair of polynomials $F(x,y)$ and $\varphi(x,y)$, an integrable
system on $\Rd$ for any $d\geq1$.
\subparg{Notation}
$\Rd$ is throughout viewed as the space of pairs of polynomials
$(u(\l),v(\l))$, with $u(\l)$ monic of degree $d$ and $v(\l)$ of degree
less than $d$, via
  $$\eqalign{u({\l})&={\l}^d+u_1{\l}^{d-1}+\cdots+u_{d-1}{\l}+u_d,\cr
               v({\l})&=\hphantom{{\l}^dpB}
               v_1{\l}^{d-1}+\cdots+v_{d-1}{\l}+v_d,}
     \formula
  $$
\formulaname{uxvx}so the coefficients $u_i$ and $v_i$ serve as
coordinates on $\Rd$. Some formulas below are simplified by denoting
$u_0=1$.
\smallskip
For any rational function $r(\l)$, we denote by
$\plus{r(\l)}$ its polynomial part and we let
$\min{r(\l)}=r(\l)-\plus{r(\l)}$. If $f(\l)$ is any polynomial and
$g(\l)$ is a monic polynomial, then $f(\l)\mod g(\l)$ denotes the
polynomial of degree less than $\deg g(\l)$, defined by
  $$ f(\l)\mod g(\l)=g(\l)\min{f(\l)\over g(\l)},
  $$
so $f(\l)=f(\l)\mod g(\l)+h(\l)g(\l)$ for a unique polynomial $h(\l)$, and
$f(\l)\mod u(\l)$ is easy computed as the rest obtained by
the Euclidean division algorithm.
\subparg{The compatible Poisson structures $\Poissonnil$}
Any polynomial $\varphi(x,y)$ specifies a Poisson bracket on $\R2$
by $ \{y,x\}=\varphi(x,y),$
which induces a polynomial bracket on the cartesian product
$\left(\R2\right)^d=\R2\times\cdots\times\R2$ by
  $$ \{y_i,x_j\}=\delta_{ij}\varphi(x_j,y_i),\qquad
      \{x_i,x_j\}=\{y_i,y_j\}=0.
    \formula
  $$
\formulaname{product_bracket}Let $\Delta$
denote the closed subsets of $\left(\R2\right)^d$  defined by
  $$ \Delta=\{((x_1,y_1),(x_2,y_2),\dots,(x_d,y_d))\mid\exists
  i,j\,\colon\, i\neq j\hbox{ and } x_i=x_j\},
  $$
and consider  the map
$\S\colon\left(\R2\right)^d\setminus\Delta\to\R{2d}$, given by
  $$\left((x_1,y_1),(x_2,y_2),\dots,(x_d,y_d)\right)\mapsto
     (u(\l),v(\l))=\left(\prod_{i=1}^d(\l-x_i),\sum_{i=1}^dy_i\prod_{j\neq
     i}{\l-x_j\over x_i-x_j}\right).
   \formula
  $$
\formulaname{map_S}$\S$ is
invariant for the obvious action of the permutation group $S_d$ on
$(\R2)^d$ and is a $d!\colon1$ covering map onto an open subset of
$\Rd$. Since the Poisson structure is also invariant for the
action of $S_d$, i.e.,
  $$ \{f,g\}\circ\sigma=\{f\circ \sigma,g\circ\sigma\},\qquad
     f,g\in C^\infty\left(\left(\R2\right)^d\right),\ \sigma\in S_d,
  $$
a $C^\infty$ Poisson bracket $\Poissonnil$ is defined on
the image of $\S$ by requiring that $\S$ is a Poisson map, i.e.,
that for any $f,g\in C^\infty(\Rd)$, one has
$\Poisson{f,g}\circ \S=\{f\circ \S,g\circ \S\}.$
The following proposition provides us with explicit formulas for this
bracket, showing in particular that it extends to
a $C^\infty$ (even polynomial) Poisson bracket on all of $\Rd$.
\Proposition{The Poisson bracket $\Poissonnil$ is given in
terms of the coordinates $u_i,v_i$ by
  $$ \eqalign{\Poisson{u({\l}),u_j}&=\Poisson{v({\l}),v_j}=0,\cr
              \Poisson{u({\l}),v_j}&=\Poisson{u_j,v({\l})}=\varphi
              (\l,v(\l))\plus{u({\l})\over {\l}^{d-j+1}}\mod u(\l),
                    \qquad 1\leq j\leq d,}
     \formula
  $$
\formulaname{Poisson}hence all brackets of the coordinate functions
$u_i$ and $v_j$ are polynomials and $\Poissonnil$
is defined on all of $\Rd$.
Except for the trivial bracket $\poissonnil^0$, all Poisson bracket
$\Poissonnil$ are of rank $2d$ on a dense subset of $\Rd$ whose
complement is a (possibly empty) algebraic hypersurface; moreover they
are all compatible, i.e., the sum of two such Poisson brackets is again
a Poisson bracket.
\smallskip
As a special and most important case, if $x$
and $y$ are canonical variables, i.e., $\varphi(x,y)=1$, then
the Poisson structure $\Poissonnil$, also denoted by
$\poissonnil$, is of maximal rank at every point of $\Rd$,
hence it defines a symplectic structure $\omega_d$ on $\Rd$; the Poisson
bracket {\rm
(6)\nameformula{Poisson}} reduces in this case to
  $$ \poisson{u(\l),v_j}=\poisson{u_j,v(\l)}=\plus{u(\l)\over
  \l^{d-j+1}},
    \formula
  $$
\formulaname{basic_Poisson}and
its matrix of Poisson
brackets with respect to the coordinate functions $u_i$ and $v_j$,
takes the form
  $$ P=\pmatrix{0& U\cr -U&0}\quad \hbox{where}\quad
      U=\pmatrix{0&0&\cdots&0&1\cr
        0&0&\cdots&1&u_1\cr
        \vdots&\vdots&&\vdots&\vdots\cr
        0&1&\cdots&u_{d-3}&u_{d-2}\cr
        1&u_1&\cdots&u_{d-2}&u_{d-1}}.
  $$
In terms of $\poissonnil$, the Poisson structure associated to
a polynomial $\varphi(x,y)$ is given by
  $$ \eqalign{\Poisson{u(\l),f}&=\varphi(\l,v(\l))\poisson{u(\l),f} \mod
                                   u(\l),\cr
              \Poisson{v(\l),f}&=\varphi(\l,v(\l))\poisson{v(\l),f} \mod
                                   u(\l).}
     \formula
  $$
\formulaname{Poisson_relation}}

\Proof
Clearly $\Poisson{u(\l),u(\mu)}=0$. The bracket $\Poisson{v(\l),v(\mu)}$
is a polynomial of degree at most $d-1$ in $\l$ and in $\mu$, which
vanishes for the $d^2$ values $(\l,\mu)=(x_i,x_j)$, where $x_i$ and $x_j$ are
roots of $u(\l)$, hence
$\Poisson{v(\l),v(\mu)}$
vanishes on the image of $\S$. If $1\leq j\leq d$, then
  $$ \eqalign{\Poisson{u_j,v(\l)}
      &=(-1)^j \Poisson{\sum_{i_1<i_2<\cdots
       <i_j}x_{i_1}x_{i_2}\cdots x_{i_j}\,,\sum_{l=1}^dy_l\prod_{k\neq l}
       {\l-x_k\over x_l-x_k}},\cr
      &=(-1)^{j-1} \sum_{i_1<i_2<\cdots
        <i_j}\sum_{t=1}^j x_{i_1}x_{i_2}\cdots \widehat{x_{i_t}}\cdots
        x_{i_j}\varphi(x_{i_t},y_{i_t})\prod_{k\neq i_t}
        {\l-x_k\over x_{i_t}-x_k},\cr
      &=(-1)^{j-1} \sum_{l\notin\{i_1<i_2<\cdots
        <i_{j-1}\}} x_{i_1}x_{i_2}\cdots
x_{i_{j-1}}\varphi(x_l,y_l)\prod_{k\neq l}
        {\l-x_k\over x_l-x_k},\cr
      &=(-1)^{j-1}\sum_{l=1}^d\varphi(x_l,y_l)\prod_{k\neq l}{\l-x_k\over
        x_l-x_k}\,(-1)^{j-1}\sum_{m=0}^{j-1}x_l^mu_{j-m-1},\cr
      &=\sum_{l=1}^d\sum_{m=0}^{j-1}x_l^mu_{j-m-1}\varphi(x_l,y_l)
        \prod_{k\neq l}{\l-x_k\over x_l-x_k}.}
  $$
\goodbreak
Since $y_l=v(x_l)$ this shows that $\Poisson{u_j,v(\l)}$ is the
(unique) polynomial in $\l$ of degree less than $d$, which takes at
$\l=x_l$ the value $\sum_{m=0}^{j-1}x_l^mu_{j-m-1}\varphi(x_l,v(x_l))$,
for $l=1,\dots,d$.
As the $x_l$ are the zeros of $u(\l)$, the same is true for
$\sum_{m=0}^{j-1}\l^mu_{j-m-1}\varphi(\l,v(\l))\mod u(\l)$,
and we find
  $$\eqalign{\Poisson{u_j,v(\l)}&=
          \sum_{m=0}^{j-1}\l^mu_{j-m-1}\varphi(\l,v(\l))\mod u(\l),\cr
         &=\varphi(\l,v(\l))\plus{u(\l)\over \l^{d-j+1}}\mod u(\l),}
  $$
which proves the second equality in
(6)\nameformula{Poisson}. For the first equality in
(6)\nameformula{Poisson}, remark that
  $$ \eqalign{\Poisson{u(\l),v(\mu)}&=\sum_{l=1}^d\Poisson{
      \prod_{i=1}^d(\l-x_i),y_l}
        \prod_{j\neq l}{\mu-x_j\over x_l-x_j},\cr
     &=\sum_{l=1}^d\varphi(x_l,y_l)\prod_{j\neq l}{(\l-x_j)(\mu-x_j)\over
     x_l-x_j},}
  $$
is symmetric in $\l$ and $\mu$, which leads at once to
$\Poisson{u(\l),v_i}=\Poisson{u_i,v(\l)}$.
\smallskip
Since $\varphi(x,y)$ is a polynomial, it follows from the fact that
$u(\l)$ is monic and formulas
(6)\nameformula{Poisson} that all brackets $\Poisson{u_i,v_j}$ are
polynomial, hence extend to a Poisson bracket on $\Rd$, also denoted by
$\Poissonnil$. Compatibility of the brackets derives from the obvious
formula
$\Poissonnil+\{\cdot\,,\cdot\}_d^\psi=\{\cdot\,,\cdot\}_d^{\varphi
+\psi}$.
\smallskip
For $\varphi=1$ one obtains
(7)\nameformula{basic_Poisson}, because the degree of $\plus{u(\l)\over
\l^{d-j+1}}$ is less than $d$ for any $j=1,\dots, d$, which also leads at
once to the matrix representation of $\poissonnil$ --- since
its determinant equals $(-1)^d$, it is of rank $2d$ everywhere.
Remark also that $\poissonnil$ is not compatible with the
standard structure $\sum du_i\wedge dv_i$ on $\Rd$.
\smallskip
To see where the rank of the Poisson structure $\Poissonnil$
fails to be maximal, we
need to investigate the determinant of the matrix of Poisson brackets
$\Poisson{u_i,v_j}$. By some elementary properties of determinants one
finds that for any values $x_1,\dots,x_d$,
  $$ \det\left(\Poisson{u_i,v(x_j)}\right)_\sub=\det\left(
     \Poisson{u_i,v_j}\right)_\sub\prod_{k<l}(x_k-x_l).
     \formula
  $$
\formulaname{det_rel}Choosing $x_1,\dots,x_d$ to be the
roots of $u(\l)$ (which may be complex), we get from
(6)\nameformula{Poisson}
  $$\eqalign{\det\left(\Poisson{u_i,v(x_j)}\right)_\sub
     &=\det\left(\varphi(x_j,v(x_j))\left[{u(\l)\over
         \l^{d-i+1}}\right]_{+\vert\l=x_j}\right)_\sub,\cr
     &=\det\left(\left[{u(\l)\over
         \l^{d-i+1}}\right]_{+\vert\l=x_j}\right)_\sub
         \prod_{m=1}^d\varphi(x_m,v(x_m)),\cr
     &=\det\left(\poisson{u_i,v(x_j)}\right)_\sub
         \prod_{m=1}^d\varphi(x_m,v(x_m)),\cr
     &{\buildrel (i)\over =}(-1)^{[d/2]}\prod_{k<l}(x_k-x_l)
         \prod_{m=1}^d\varphi(x_m,v(x_m)),\cr}
  $$
where in {\it (i)} we used
(9)\nameformula{det_rel} for $\varphi=1$. It follows that (even if $u(\l)$
has multiple roots)
  $$ \det\left(\Poisson{u_i,v_j}\right)_\sub=(-1)^{[d/2]}
         \prod_{m=1}^d\varphi(x_m,v(x_m)),
  $$
on all of $\Rd$, hence the Poisson structure is of lower rank on the locus
$\prod_{j=1}^d\varphi(x_j,v(x_j))=0$, which for given $\varphi$ is
easy written as the equation of an algebraic hypersurface in $\Rd$.
\smallskip
Finally,
(8)\nameformula{Poisson_relation} follows immediately from the Leibniz
property of Poisson brackets.
\endproof
\Amplification
The condition that $\varphi(x,y)$ is a polynomial is not essential: if
$\varphi(x,y)$ is any smooth function,
then then all the above formulas remain valid,
yielding yet more examples of compatible Poisson structures.
In this more general
case, for $f(\l)$ any smooth funtion and $g(\l)$ a monic polynomial as
before, $f(\l)\mod g(\l)$ denotes the unique\footnote{1}{If
$g(\l)$ has multiple roots, then $f(\l)\mod g(\l)$ is not unique; since
in this paper $g(\l)=u(\l)$ depends on the coordinates $u_i$, it is (as
a function on $\Rd$) uniquely defined on a dense subset of $\Rd$, hence
its extension to $\Rd$ is also unique.} polynomial of degree less than
$\deg g(\l)$ which takes at the roots $x_i$ of $g(\l)$ the value
$f(x_i)$. The Poisson brackets $\Poisson{u_i,v_j}$ are no longer
polynomial and can not be computed by the Euclidean division algorithm.

Of interest is also the case that $\varphi(x,y)$ is rational, in which
all brackets $\Poisson{u_i,v_j}$ are rational functions of the
coordinates $u_i$ and $v_j$. Obviously, if $\varphi(x,y)$ has poles on
$\R2$, the bracket $\Poissonnil$ will also have poles on $\Rd$, and is
in this case only
a Poisson bracket on a
dense subset of $\Rd$.
\subparg{Polynomials in involution for $\Poissonnil$}
We now show how there is associated, for fixed $d$, to each polynomial
$F(x,y)$ which depends explicitely on $y$, a set of $d$
independent polynomials, which are in involution for all the brackets
$\Poissonnil$, that is, the Poisson bracket of any pair of
these polynomials vanishes (such polynomials are also said to {\it
Poisson-commute\/}).
\medskip
Let $F(x,y)\in\Real[x,y]\setminus\Real[x]$ and let us
view $\R d$ as the space of polynomials (say in
$\l$) of degree less than $d$. Then there is a natural map $\hat H_{F,d}$
from
$\left(\R2\right)^d\setminus\Delta$ to $\R d$, which assigns to a
$d$-tuple $((x_1,y_1),\dots,(x_d,y_d))$ the unique polynomial in
$\Real[\l]$ of degree
less than $d$, which takes for  $\l=x_i$ the value $F(x_i,y_i)$ (for
$i=1,\dots,d$). Since $\hat H_{F,d}$ is invariant under the action of $S_d$,
there is defined on the image of $\S$ a map $H_{F,d}$ such that $\hat
H_{F,d}=H_{F,d}\circ
\S$, namely $H_{F,d}$ is  given by
  $$ H_{F,d}(u(\l),v(\l))=F({\l},v({\l})) \mod u({\l}).
     \formula
  $$
\formulaname{H}The $d$ components of the
map $H_{F,d}$ define $d$ functions on $\Rd$, which
will be simply denoted by $H_1,\dots,H_d$ (omitting the dependence on
$F$ and $d$ in the notation), i.e.,
$H_{F,d}(u(\l),v(\l))=H_1\l^{d-1}+H_2\l^{d-2}+\cdots+H_d$.
As $u({\l})$ is a monic polynomial,
these functions $H_i$ are  polynomial in our
coordinates on $\Rd$ hence are defined on all of $\Rd$.
The main result of this section is the following.
\Theorem{The coefficients $H_1,\dots,H_d$ of
$H({\l})=F({\l},v({\l}))\mod u({\l})$ define
for any non-zero polynomial $\varphi(x,y)$ a completely integrable system
on the Poisson manifold $\left(\Rd,\Poissonnil\right)$
with polynomial invariants, that
is, $\{H_1,\dots,H_d\}$ forms a set of $d$ functional independent polynomials
on $\Rd$,
which are in involution for all brackets $\Poissonnil$.}
\propositionname{main}
\smallskip
Before proving this theorem we prove a key lemma and
write down explicit equations for the
Hamiltonian vector fields $X^\varphi_{H_i}=\Poisson{\cdot,H_i}$, which
--- by the above theorem ---
commute as differential operators, in view of the identity (see
\referentie{AM}) $\left[X^\varphi_{H_i},X^\varphi_{H_j}\right]=
        X^\varphi_{\Poisson{H_j,H_i}}.$
\Lemma{Let $p({\l}),\,q({\l})$ and $r(\l)$ be polynomials, with $\deg
q(\l)\geq\deg r(\l)$ and let $i\in\N$.
  $$ \eqalign{&{\it (1)} \qquad
      r({\l})\plus{\l^{-i}q({\l})}\mod q(\l)=
      r({\l})\plus{\l^{-i}q({\l})}-q({\l})\plus{\l^{-i}r({\l})},\cr
      &{\it (2)} \qquad
      \sum_{l=1}^{\deg q}\m^{l-1}p({\l})\plus{\l^{-l}q({\l})}\mod q({\l})=
    \sum_{l=1}^{\deg q}{\l}^{l-1}p(\m)\plus{\m^{-l}q(\m)}\mod q(\m).}
      \formula
  $$
\formulaname{exchange}}
\lemmaname{exchange_lemma}
\Proof
For the proof of {\it (1)\/} remark that if
$\deg r({\l})
\leq\deg q({\l})$ then the right hand side of the identity
  $$ r({\l})\plus{\l^{-i}}-q({\l})\plus {\l^{-i}r({\l})}=
               -r({\l})\min {\l^{-i}q({\l})}+q({\l})\min
               {\l^{-i}r({\l})}
  $$
is of degree less than $\deg q({\l})$, hence also the left hand side.
To show
{\it (2)\/} we may assume that $\deg p({\l})<\deg q({\l})$ because
the identity depends only
on $p({\l})\mod q({\l})$. Then
  $$\eqalign{\sum_{l=1}^{\deg q}{\l}^{l-1}p(\m)\plus{\m^{-l}q(\m)}\mod q(\m)
                                 &{\buildrel (i)\over=}
   \sum_{l=1}^{\deg q}{\l}^{l-1}\left(p(\m)\plus{\m^{-l}q(\m)}
                -q(\m)\plus{\m^{-l}p(\m)}\right),\cr
                        &{\buildrel (ii)\over=}
                   \sum_{l=1}^{\deg
                   q}\m^{l-1}\left(p({\l})\plus{\l^{-l}q({\l})}
                -q({\l})\plus{\l^{-l}p({\l})}\right),\cr
        &=\sum_{l=1}^{\deg q}\m^{l-1} p({\l})\plus{\l^{-l}q({\l})}
                \mod q({\l}).}
  $$
In {\it (i)\/} we applied part {\it (1)\/} of this lemma;
the exchange property in {\it (ii)} is proven at
once by expanding the polynomials or by induction on $\deg q(\l)$.
\endproof
\Proposition{The coefficients $H_i$ of $F(\l,v(\l))\mod u(\l)$
determine $d$ independent polynomial
vector fields $X^\varphi_{H_i}$ on $\Rd$,
which are explicitely given by
  $$ \eqalign{X^\varphi_{H_i}u({\l})
         &=\varphi(\l,v(\l)){\partial F\over\partial
y}({\l},v({\l}))\plus{u({\l})\over
        {\l}^{d-i+1}}\mod u({\l}),\cr
        X^\varphi_{H_i}v({\l})&=\varphi(\l,v(\l))
           \plus{F({\l},v({\l}))\over u({\l})}\plus{u({\l})\over
                {\l}^{d-i+1}}\mod u({\l}).}
     \formula
  $$
\formulaname{vector_field}Moreover, the following remarkable identities hold
for all $1\leq i,j\leq d$:
  $$\Poisson{u_i,H_j}=\Poisson{u_j,H_i}\hbox{ and }
             \Poisson{v_i,H_j}=\Poisson{v_j,H_i}.
                 \formula
  $$
\formulaname{identities}}
\Proof
Writing $X_{H_i}$ as a shorthand for $X^1_{H_i}$,
we first compute $X_{H_i}u({\l})=\poisson{u({\l}),H_i}$, which we obtain
as the
coefficients of $\m^{d-i}$ in $\poisson{u({\l}),H_{F,d}(u(\m),v(\m))}$.
  $$\eqalign{\poisson{u({\l}),H_{F,d}(u(\m),v(\m))}
              &=\sum_{j=1}^d\poisson{u({\l}),v_j}
               {\partial H_{F,d}\over\partial v_j}(u(\m),v(\m)),\cr
              &=\sum_{j=1}^d\plus{u({\l})\over {\l}^{d-j+1}}{\partial
              H_{F,d}\over
                \partial v_j}(u(\m),v(\m)),\cr
              &=\sum_{j=1}^d\sum_{k=0}^{j-1} u_k {\l}^{j-k-1}
              {\partial F\over\partial y}(\m,v(\m))\m^{d-j}\mod u(\m),\cr
              &=\sum_{l=1}^{d}\sum_{j=l}^d u_{j-l} {\l}^{l-1}
                     {\partial F\over\partial y}(\m,v(\m))\m^{d-j}
                                                 \mod u(\m),\cr
              &=\sum_{l=1}^{d}{\l}^{l-1}{\partial F\over\partial y}
                        (\m,v(\m))
                           \plus{u(\m)\over\m^{l}}\mod u(\m),\cr
              &=\sum_{l=1}^{d}\m^{l-1}{\partial F\over\partial y}({\l},v({\l}))
              \plus{u({\l})\over {\l}^{l}}\mod u({\l}),}
  $$
where we used the exchange property
(11)\nameformula{exchange} in the last step.
Since $H_i$ is the coefficient of $\m^{d-i}$ in $H({\l})$
this gives equation
(12)\nameformula{vector_field} for $X_{H_i}u({\l})$ in case $\varphi(x,y)=1$.
In a similar way
$X_{H_i}v({\l})$ is found, the computation of ${\partial \over\partial u_j}
H_{F,d}(u(\m),v(\m))$ is however more involved: let $1\leq j\leq d$ then
  $$ \eqalign{{\partial\over\partial u_j}\left(F(\m,v(\m))\mod u(\m)\right)
                                  &={\partial\over\partial u_j}\left(u(\m)
               \min{F(\m,v(\m))\over
                 u(\m)}\right),\cr
                &=-{\partial\over\partial u_j}\left(u(\m)\plus{F(\m,v(\m))\over
                u(\m)}\right),\cr
                &=-u(\m)\left({\m^{d-j}\over u(\m)}\plus{F(\m,v(\m))\over
u(\m)}
                -\plus{{\m^{d-j}\over u(\m)}{F(\m,v(\m))\over u(\m)}}
                \right),\cr
                &{\buildrel (i)\over =}-u(\m)\min{{\m^{d-j}\over u(\m)}
                \plus{F(\m,v(\m))\over u(\m)}},\cr
                &=-\m^{d-j}\plus
                {F(\m,v(\m))\over u(\m)}\mod u(\m).}
  $$
In {\it (i)} we used that if $R=R(\m)$ and $P=P(\m)$ are rational functions,
with $\plus{R}=0$, then
  $$ R\plus{P}-\plus{RP}=R\plus{P}-\plus{R\plus{P}}=\min{R\plus{P}}.
  $$
Granted this we obtain as above
  $$\eqalign{\poisson{v({\l}),H_{F,d}(u(\m),v(\m))}
                &=\sum_{j=1}^d\m^{d-j}\plus{u({\l})\over
{\l}^{d-j+1}}\plus{F(\m,v(\m))
                \over u(\m)}\mod u(\m),\cr
                &=\sum_{l=1}^{d}\m^{l-1}\plus{u({\l})\over
                {\l}^{l}}\plus{F(\l,v(\l))\over u(\l)}\mod u({\l}),}
  $$
which leads at once to the expression
(12)\nameformula{vector_field} for $X_{H_i}v({\l})$ in case
$\varphi(x,y)=1$. Having obtained the formulas
(12)\nameformula{vector_field}
for $X_{H_i}u(\l)$ and $X_{H_i}v(\l)$, the
formulas for $X^\varphi_{H_i}u(\l)$ and $X^\varphi_{H_i}v(\l)$,
are obtained at once upon using
(8)\nameformula{Poisson_relation}.
\medskip
Finally, the exchange property
(11)\nameformula{exchange} implies that
${\l}$ and $\m$ are everywhere interchangeable in the
above computations so we get $\Poisson{u({\l}),H_{F,d}(u(\m),v(\m))}
=\Poisson{u(\m),H_{F,d}(u({\l}),v(\l))}$, which is tantamount
to the identity $\Poisson{u_i,H_j}=\Poisson{u_j,H_i}$. The second formula in
(13)\nameformula{identities} follows in the same way.

\endproof

\Proofof{Theorem
2\nameproposition{main}}
We first prove
that $\Poisson{H_i,H_{F,d}(u(\l),v(\l))}=0$ for $1\leq i\leq d$. To make the
proof
more transparent, we use the following abbreviations:
  $$ F_y={\partial F\over\partial y}({\l},v({\l})),\quad
       F_{(u)}={F({\l},v({\l}))\over u({\l})}\quad\hbox{and}\quad
            U_i={\varphi(\l,v(\l))\over u({\l})}\plus{u({\l})\over
            {\l}^{d-i+1}},
  $$
so that
(12)\nameformula{vector_field} is rewritten as
$X^\varphi_{H_i}u({\l})=u({\l})\min{U_iF_y}$ and
$X^\varphi_{H_i}v({\l})=u({\l})\min{U_i\plus{F_{(u)}}}$. Then
  $$\eqalign{\Poisson{H_{F,d}(u(\l),v(\l)),H_i}
        &=X^\varphi_{H_i}\left(u({\l})\min{F({\l},v({\l}))\over
u({\l})}\right),\cr
        &=X^\varphi_{H_i}u({\l})\min{\Fu}+u({\l})\min{X^\varphi_{H_i}
          F({\l},v({\l}))\over u({\l})}
        -u({\l})\min{F_{(u)}X^\varphi_{H_i}u({\l})\over u({\l})},\cr
        &=u({\l})\left(\min{U_iF_y}\min{\Fu}+\min{F_y\min{U_i\plus{\Fu}}}
        -\min{\Fu\min{U_iF_y}}\right),\cr
        &=u({\l})\min{\min{U_iF_y}\min{\Fu}+F_y\min{U_i\plus{\Fu}}
        -\Fu\min{U_iF_y}},\cr
        &{\buildrel (i)\over =} u({\l})\min{-\min{U_iF_y}\plus{\Fu}+
        F_yU_i\plus{\Fu}},\cr
        &=u({\l})\min{\plus{U_iF_y}\plus{\Fu}},\cr
        &=0.}
  $$
In {\it (i)} we used the fact that $F_y$ is a polynomial, i.e., $\min{F_y}=0$.
\smallskip
We now show that the $d$ coefficients of
$H_{F,d}(u(\l),v(\l))=F({\l},v({\l}))\mod u({\l})$ are
functional independent. Clearly the last $d$ coefficients $\tilde H_1,\dots,
\tilde H_d$ of $F({\l},v({\l}))$ are independent because $v_i$ appears only in
$\tilde H_1,\dots,\tilde H_i$ (it {\it does} appear since
$F(x,y)\notin\Real[x]$).
Reducing $F({\l},v({\l}))$ modulo $u({\l})$
amounts to substracting from $\tilde H_i$ polynomials of lower degree in
the variables $v_j$, so it cannot make these functions dependent and the
independence of $\{H_1,\dots,H_d\}$ follows.
\endproof
\medskip
\Amplification
If $F(x,y)$ and $F'(x,y)$ differ only by a polynomial
which is independent of $y$ and is of degree less than $d$ in $x$, then
clearly the $d$-dimensional integrable systems
which are associated to $F$ and $F'$ are
the same; in this sense, for $\varphi(x,y)$ fixed,
a system is associated to a coset
  $$ \tilde F(x,y)=\left\{F(x,y)+\sum_{i=0}^{d-1}c_ix^i\mid c_i\in \Real
  \right\}.
  $$
If a (differentiable) deformation family $M$ of classes
$\tilde F(x,y)$ is given (rather than a single class) then our
construction is easy adapted to give (for each non-zero
$\varphi(x,y)\in\Real[x,y]$) a $d$-dimensional
integrable system on a Poisson
manifold, which is the product of the deformation manifold $M$ and
$\Rd$. Namely let the brackets
(4)\nameformula{product_bracket} on $\left(\R2\right)^d$ be extended
trivially to $\left(\R2\right)^d\times M$, i.e., if $\pi_M$ denotes the
projection map $\left(\R2\right)^d\to M$ then the annihilator of the
Poisson bracket is chosen as $\left\{f\circ\pi_M\mid f\in
C^\infty(M)\right\}$. Also the map $\S$ given by
(5)\nameformula{map_S} is extended to the map
  $$ \S\times\Id_M\colon\left(\left(\R2\right)^d\setminus\Delta\right)\times
        M\to\Rd\times M,
  $$
which is the identity  map $\Id_M$ on the second component. As both
this Poisson structure and these maps are again invariant for the action
of $S_d$ (on the first component) we obtain a Poisson structure
$\{\cdot\,,\cdot\}_{d,M}^\varphi$ on the image of
$\S\times\Id_M\subset\Rd\times M$, which extends to all of $\Rd\times
M$, again because all brackets  are polynomial.
The commuting vector fields
$\{\cdot,H_i\}_{d,M}^\varphi$ are tangent to the (linear)
Poisson submanifolds
$\{\tilde F\}\times\Rd,\,(\tilde F\in M)$, to which
$\{\cdot\,,\cdot\}_{d,M}^\varphi$ restricts as
$\{\cdot\,,\cdot\}_{d}^\varphi$. Therefore, these commuting vector
fields  restricts to these submanifolds and
give there the vector fields $\{\cdot,H_i\}^\varphi_d$
(as given by
(12)\nameformula{vector_field}) of the integrable system
associated to $\tilde F$ (i.e., to $F$).
\medskip
\Amplification
In all the above definitions, $\Real$ can be replaced
by $\Complex$; our construction then associates to
each complex polynomial in two variables, a maximal set of holomorphic
functions (polynomials), defined on $\C{2d}$, which are in involution
with respect to a holomorphic Poisson bracket, itself determined by an
arbitrary non-zero polynomial in two variables.
\endparg
\parg{The geometry of the invariant manifolds}
\sectionname{invariant_manifolds}The
integrable systems introduced in Section
2\namesection{systems} provide us (for each $d\geq1$ and
$F(x,y)\in\Real[x,y]\setminus\Real[x]$) with a surjective
map
defined by $H_{F,d}(u(\l),v(\l))=F({\l},v({\l}))\mod u({\l})$.
The fibers of $H_{F,d}$ are preserved by the flows
of the $d$
vector fields $X^\varphi_{H_i}$ which correspond via $\Poissonnil$ to the
components of
this map. By Sard's Theorem, the generic fiber of this map is smooth. These
smooth fibers are called the {\it invariant manifolds} of the system; they are
Lagrangian submanifolds of $(\Rd,\Poissonnil)$, i.e., the restriction of
$\Poissonnil$ to these $d$-dimensional submanifolds vanishes.
In this section we investigate the geometry of these invariant
manifolds and discuss their compactification.
\subparg{The invariant manifolds $\AFd$ and {\rm $\AFdC$}}
Since $H_{F,d}(u(\l),v(\l))$ is defined as $F({\l},v({\l}))\mod u({\l})$, the
fiber
over $h({\l})\in
\Real_{d-1}[\l]$ is the same as the fiber over $0$ for
$H_{F',d}$, where
$F'(x,y)=F(x,y)-h(x).$ Therefore we may restrict ourselves
to the fiber
lying over $0$, denoted  by $\AFd$; thus, by definition, $\AFd$ is given by
  $$\AFd=\left\{(u(\l),v(\l))\in\Rd\mid
                 \min{F({\l},v({\l}))\over u({\l})}=0\right\}.
       \formula
  $$
\formulaname{AFd}Sard's Theorem
says now that this fiber is smooth
if $F(x,y)$ is generic. Clearly if $F(x,y)$ is generic then the
{\sl complex\/} algebraic curve $\GF\subset\C2$, defined by $F(x,y)=0$,
is smooth. We show now that in fact the latter suffices for $\AFd$ to be
smooth.
\Proposition{If the  algebraic curve $\GF\subset\C2$
defined by $F(x,y)=0$ is smooth, then the fiber $\AFd\subset\Rd$ is also
smooth.}
\propositionname{smooth_fiber}\Proof
$\AFd$ will be smooth if and only if $H_{F,d}$ is submersive at
each point of $\AFd$, i.e., iff
  $$\hbox{rank}\left({\partial H_i\over\partial u_1},\dots,{\partial H_i\over
            \partial u_d},{\partial H_i\over\partial v_1},\dots,
                {\partial H_i\over \partial v_d}\right)_{1\leq i\leq d}=d
                \hbox{ along } \AFd.
  $$
{}From the proof of Theorem
2\nameproposition{main} and the definition
(14)\nameformula{AFd} of $\AFd$, the $j$-th and $d+j$-th columns of this
matrix are respectively given by
  $${\l}^{d-j}{F({\l},v({\l}))\over u({\l})}\mod u({\l})\quad\hbox{and}
         \quad {\l}^{d-j} {\partial F\over
         \partial y}({\l},v({\l}))\mod u({\l}).
  $$
It is therefore sufficient to show that if $\GF$ is smooth then
the dimension of the linear space
  $$\left(R_1({\l}){F({\l},v({\l}))\over u({\l})}+R_2({\l})
         {\partial F\over\partial y}({\l},v({\l}))
        \right)\mod u({\l}),\ \deg R_i({\l})<d,
    \formula
  $$
\formulaname{R1R2}equals $d$. Let ${\l}_1,\dots,{\l}_r$ be the distinct roots
of $u({\l}),\,{\l}_i$ having multiplicity $s_i$. We claim that
  $${F({\l}_i,v({\l}_i))\over u({\l}_i)}=0 \quad\hbox{and}
              \quad{\partial F\over\partial y}
         ({\l}_i,v({\l}_i))=0
     \formula
  $$
\formulaname{singular}cannot hold simultaneously if $\GF$ is smooth.
For otherwise $(x_i,y_i)=
(\l_i,v(\l_i))$ would be a singular point of $\GF$: if
(16)\nameformula{singular} holds then clearly ${\partial F
\over\partial y}(x_i,y_i)=0$, but also $F(x_i,y_i)=
{\partial F\over\partial x}(x_i,y_i)=0$
because in this case $F(x,y_i)$ has a double zero at $x=x_i$.
\smallskip
The dimension of
(15)\nameformula{R1R2} is now investigated by remarking that
for any polynomial $p({\l})$, the
value of $p({\l})\mod u({\l})$ at ${\l}_i$ is just $p({\l}_i)$, and the
values of the first $s_i-1$
derivatives of $p({\l})\mod u({\l})$ at $\l_i$ are given by the values of the
corresponding
derivatives of $p(\l)$ at ${\l}_i$.
Let us suppose that the different roots of $u(\l)$ are ordered such that
${\l}_1,\dots,{\l}_t$ are also zeros of
${\partial F\over\partial y}(\l,v(\l))$,
while ${\l}_{t+1},\dots,{\l}_r$ are not. As a first
restriction, let $R_1({\l})$ (resp.\ $R_2({\l})$) be such that its first
$s_i-1$
derivatives vanish at ${\l}_i$ for $t+1\leq i\leq r$ (resp.\ $1\leq i\leq t$).
As a further restriction it is (by the first restriction and as
(16)\nameformula{singular} cannot happen) now easy to see that
$R_1({\l})$ (resp.\ $R_2({\l})$) can be determined such that the
polynomial given by
(15)\nameformula{R1R2} and the first $s_i-1$ derivatives of
(15)\nameformula{R1R2}
take any given values at
${\l}_i$ for $1\leq i\leq t$ (resp.\ $t+1\leq i\leq r$). These $d$
conditions are independent, hence the dimension of
(15)\nameformula{R1R2} equals $d$ and $\AFd$ is smooth.
\endproof
We aim at a more precise description of the structure of
the invariant manifolds $\AFd$, which will be useful for describing
their topological structure.
If the fixed point set of the complex conjugation map $\tau\colon\C {2d}\to
\C {2d}\colon z\mapsto\bar z$ is denoted as $\hbox{Fix}(\tau)$, then clearly
$\AFd$ is given by
  $$\AFd=\hbox{Fix}(\tau)\cap\AFd^{\Complex},
       \hbox{ where }
  \AFd^{\Complex}=\left\{(u(\l),v(\l))\in\C{2d}\mid
                 \min{F({\l},v({\l}))\over u({\l})}=0\right\}.
       \formula
  $$
\formulaname{AFdC}Therefore, $\AFd$ is called a {\it real} algebraic variety
(see \referentie{S}). Remark that $\AFdC$ is the complex invariant manifold
lying over $0$ of the integrable system on $\C {2d}$ associated to $F$
(see Amplification 3). The following proposition is the complex analog
of Proposition
5\nameproposition{smooth_fiber}.
\Proposition{The curve $\GF\subset\C2$ is smooth if and only if the
fiber $\AFdC\subset\C{2d}$ is smooth.}
\Proof
If $\GF$ has a
singular point $P_1=(x_1,y_1)$, choose $d-1$ different points
$P_i=(x_i,y_i)$ on $\GF$ and define
$(u(\l),v(\l))=\S((x_1,y_1),\dots,(x_d,y_d))\in\AFdC$. All polynomials given by
(15)\nameformula{R1R2} vanish for $\l=x_1$, hence they span a linear space
of dimension less than $d$. Thus $H_{F,d}$ is not submersive at
$(u(\l),v(\l))$ and $\A_{F,d}$ is singular at this point.
This shows the if part of the proposition; the only if part is proven
verbatim as in the real case (Proposition
5\nameproposition{smooth_fiber}).
\endproof
It will be seen that a clear understanding of the structure of the
complex manifolds $\AFdC$ (for $\GF$ smooth), leads also to a precise
description of the real manifolds $\AFd$.
\subparg{The structure of the complex invariant manifolds {\rm $\AFdC$}}
We will show that $\AFd^{\Complex}$ is an affine part of the $d$-fold
symmetric product $\Symd\GF$ of $\GF\subset\R2$. Recall (e.g.\ from
\referentie{Gu}) that $\Symd\GF$ is defined as the orbit space of the
obvious action of the permutation group $S_d$ on the cartesian
product $\GF^d=\GF\times\cdots\times\GF$ ($d$ factors), i.e.,
  $$\Symd\GF=\GF^d/S_d.
  $$
$\Symd\GF$ inherits its structure as a complex
algebraic variety from the algebraic structure of $\GF$. Moreover the
smoothnes of $\GF$ implies smoothnes of $\Symd\GF$: namely each point
$P=\langle P_1^{m_1},\dots,P_r^{m_r}\rangle\in\Symd\Gamma$ (with all $P_i$
different; $m_i$ is the multiplicity of $P_i$ in $P$) has a neighborhood which
is isomorphic to a neighborhood of $\left(\langle P_1^{m_1}\rangle,
\dots,\langle P_r^{m_r}\rangle\right)$
in $\Sym{m_1}\GF\times\cdots\times\Sym{m_r}\GF$, and a point
$\langle  P_i^{m_i}\rangle$ on the diagonal of $\Sym{m_i}\GF$ has coordinates
given by the $m_i$ elementary symmetric functions of the $m_i$ coordinate
functions on $\GF^{m_i}$.
\goodbreak
\Theorem{If the algebraic curve $\GF$ in $\C2$, defined by $F(x,y)=0$
is smooth, then $\AFdC$ is biholomorphic to the (Zariski) open subset of
$\Symd\GF$, obtained by removing from it the divisor
  $$ \DFd=\left\{\langle P_1,\dots,P_d\rangle\mid\exists i,j\colon
          1\leq i< j\leq d,
          \left(\eqalign{&x(P_i)=x(P_j)\hbox{ with } P_i\neq P_j
              \hbox{, or }\cr
              &P_i=P_j \hbox{ is a ramification point of } x
              }\right)\right\}.
  $$}
\propositionname{biholomorphism}
\Proof
\medskip\noindent
$\bullet\quad$ {\it Construction of the map $\phi_{F,d}\colon\AFdC\to
\Symd\GF\setminus\DFd$}
\smallskip\par\noindent
Given a point $(u(\l),v(\l))\in\AFd^{\Complex}$,
a point in $\Symd\GF$ is associated
to it as follows: for every root $\l_i$ of $u(\l)$ one has
$F(\l_i,v(\l_i))=0$, because $\min{F(\l,v(\l))\over u(\l)}=0$, so each root
$\l_i$ of $u(\l)$ determines a point $(\l_i,v(\l_i))$ on $\GF$.
Thus there corresponds to $(u(\l),v(\l))\in\AFdC$ an unordered set of $d$
points
$\langle P_1,\dots,P_d\rangle\in\Symd\GF$, where $P_i$ is defined by
$(x(P_i),y(P_i))=(\l_i,v(\l_i))$. Clearly, if $x(P_i)=x(P_j)$ then
$P_i=P_j$; therefore, to show that $\langle P_1,\dots,P_d\rangle$ stays away
from $\DFd$ we only need to prove that $P_i=P_j$ cannot occur for $i\neq
j$ if $P_i$ is a ramification point for $x$, i.e., if $y(P_i)$ is a
multiple root of $F(x(P_i),y)$ (as a polynomial in $y$). As
$P_i=P_j\,(i\neq j)$ implies that $u(\l)$ has a multiple root $x(P_i)$,
in such a case $F(x,y(P_i))$ would have a multiple root $x=x(P_i)$,
again because $\min{F(\l,v(\l))\over u(\l)}=0.$ If moreover $P_i$ is
a ramification point of $x$ then also ${\partial F\over\partial
y}(x(P_i),y(P_i))=0$ and it follows that $(x(P_i),y(P_i))$ is a
singular point of $\GF$, a contradiction.
\medskip\noindent
$\bullet\quad${\it $\DFd$ is a divisor on $\Symd\GF$}
\smallskip\par\noindent
This means that $\DFd$ is given locally as the zero locus of a
holomorphic function. If $\langle P_1,\dots,P_g\rangle\in\DFd$ let the
set of indices $\{1,\dots,d\}$ be decomposed as $S_1\cup\cdots\cup S_n$,
such that all points $P_i$ where $i$ runs through one of the subsets $S_j$
have the same $x$-coordinate, which is disjoint from the $x$-coordinates
of the points which correspond to the other subsets. For each
$P_i\,(i=1,\dots,d)$ let
$x_i$ denote the lifting of $x$ to a small neighborhood of $\langle
P_1,\dots,P_d\rangle$ (corresponding to the factor $P_i$). Then
a local defining equation of $\DFd$ is given by
  $$ \prod_{i=1}^n\prod_{j,k\in S_i\atop j<k} (x_j-x_k)=0.
  $$
\medskip\noindent
$\bullet\quad${\it $\phi_{F,d}$ is a biholomorphism}
\smallskip\par\noindent
We first construct the inverse of $\phi_{F,d}$, which is closely related
to the map $\S$, as given by
(5)\nameformula{map_S}. Let $\langle
P_1,\dots,P_d\rangle\in\Symd\GF\setminus\DFd$. Clearly $u(\l)$ is taken
as
  $$ u(\l)=\prod_{i=1}^d(\l-x(P_i)).
     \formula
  $$
\formulaname{u}If all $x(P_i)$ are different then $v(\l)$ is uniquely
determined as the
polynomial of degree $d-1$ whose value at $\l=x(P_i)$ is $y(P_i)$, i.e.,
$v(\l)$ is given by
  $$ v(\l)=\sum_{l=1}^dy_l\prod_{k\neq l}{\l-x_k\over x_l-x_k}
     \formula
  $$
\formulaname{v}and is holomorphic there. If two values coincide, say
$x(P_1)=x(P_2)$,
then $P_1=P_2$ is not a ramification point (since we stay away from
$\DFd$), hence the equation $F(x,y)=0$ can be solved uniquely as
$y=f(x)$ in a neighborhood of $P_1=P_2$. For $P_1'$ and $P_2'$ in this
neighborhood, substitute
  $$ f(x(P_i'))=f(x(P_1))+\left(x(P_i')-x(P_1)\right){df\over
      dx}(x(P_1))+{\cal O}\left(x(P_1')-x(P_1)\right)^2,\qquad (i=1,2)
  $$
for $y_1$ and $y_2$ in
(19)\nameformula{v}, to obtain that $v(\l)$ has no poles as $P_1',P_2'\to
P_1$, hence extends to a holomorphic function on the larger subset where
at most two points coincide. Since the complement of this larger subset in
$\Symd\GF\setminus\DFd$ is of codimension at least two, $v(\l)$ extends to a
holomorphic function on $\Symd\GF\setminus\DFd$. It also follows that
this holomorphic function is the inverse of $\phi_{F,d}$ on all of
$\Symd\GF\setminus\DFd$: if the point $P_i$ has multiplicity $s_i$, then
the first $s_i-1$ derivatives of $v(\l)$ at $x(P_i)$ coincide with those
of $f(\l)$ at $x(P_i)$, hence $F(\l,y(P_i))$ has a zero of order $s_i$
at $\l=x(P_i)$. Finally, the inverse of a holomorphic bijection between
complex manifolds is
always holomorphic (see \referentie{GH}), hence $\phi_{F,d}$ is a
biholomorphism.
\endproof
\subparg{The structure of the real invariant manifolds $\AFd$}
Since $\AFd$ is given as $\AFdC\cap\hbox{Fix}(\tau)$, it consists of
those polynomials $(u(\l),v(\l))\in\AFd$ whose coefficients
are all real. We figure out
what this means for the corresponding point in $\Symd\GF$.
\Proposition{Under the biholomorphism $\phi_{F,d}$, the real invariant
manifolds $\AFd$ correspond to the set of all unordered $d$-tuples of
points $\langle P_1,\dots,P_d\rangle$ on $\GF$, consisting only of real
points $P_i\in\R2\cap\GF$ and complex conjugated pairs $P_i=\bar P_j$,
each ramification point (of $x$) occurring at most once, and
$x(P_i)=x(P_j)$ only if $P_i=P_j$. Moreover its manifold structure
derives from the structure of the $d$-fold symmetric product of $\GF$.}
\propositionname{real}
\Proof
$u(\l)$ is real if and only if
its roots consist only of real roots and roots which
occur in complex conjugate pairs. Obviously, if
$v(\l)$ is real, then at each
root $x_i$ of $u(\l)$, with multiplicity $s_i$, $v(\l)$ and the first
$s_i-1$ derivatives of $v(\l)$ take complex conjugate values when
evaluated at complex conjugate points (in particular, real values at real
points). It is checked that this is also a sufficient condition for
$v(\l)$ to be real. Since $v(x_i)=y_i$, this means that the real
polynomials $(u(\l),v(\l))$ on $\AFdC$ correspond to those points
$\langle P_1,\dots,P_d\rangle$ in $\Symd\GF$ consisting of
real points $P_i=(x(P_i),y(P_i))\in\R2$ and complex conjugated pairs
$P_j=(x(P_j),y(P_j))=\left(\overline{x(P_k)},
\overline{y(P_k)}\right)=\bar P_k$, but not belonging to
$\DFd$, i.e.,
the multiplicity of each ramification point (of $x$) is at most one, and
$x(P_i)=x(P_j)$ only if $P_i=P_j$.
\endproof
Proposition
8\nameproposition{real} can be used to obtain a precise description of
the topology of the real invariant manifolds $\AFd$, as we show now for
$d=2$ (for $d=1$, $\AFd$ is just $\GF\cap\R2$, the real part of $\GF$).
For a fixed $F$ such that $\GF$ is smooth,
let the connected components of $\GF\cap\R2$ (if any) be denoted by
$\G1,\dots,\G s$ and define for $1\leq i,j,k\leq s,\,i<j$
  $$\eqalign{\G{00}&=\{\langle P,\bar P\rangle\mid
P\in\GF,\,x(P))\notin\Real\},\cr
             \G{ij}&=\{(P_1,P_2)\in\G i\times\G j\mid x(P_1)=x(P_2)
             \Rightarrow P_1=P_2\},\cr
             \G{kk}&=\left\{\langle P_1,P_2\rangle\in{\G k\times\G k\over
             S_2}\mid x(P_1)=x(P_2)\Rightarrow (P_1=P_2\hbox{ and
             is not a ramification point of $x$})\right\}}
  $$
Then the union of $\G{00}$ with all the sets $\G{ij}$ and $\G{kk}$ is
easy identified with $\phi_{F,2}(\A_{F,2})$, the surface to be described. It is
remarked
that the only paths in it which are not contained in $\R2$, are in
$\G{00}$, and $\G{00}$ connects exactly the surfaces $\G{kk}$. Therefore,
if $i\neq j$ then $\G{ij}$ is not connected to any other
$\G{mn},\,\G{kk}$, nor to $\G{00}$.
\smallskip
Therefore we first concentrate on such a subset $\G{ij}$,
say on $\G{12}$. If the
intervals $x(\G1)$ and $x(\G2)$ are disjoint, then
$\G{12}=\G1\times\G2$, so $\G{12}$ is either homeomorphic to
a torus, a cylinder or a disc,
depending on whether the components
$\G1$ and $\G2$ are closed or open. If $x(\G1)$ and
$x(\G2)$ have a point $P$ in common, then one finds again these surfaces,
but with a number of punctures (holes), equal to
  $$ \prod_{i=1}^2\#\{Q\in\G i\mid x(Q)=P\}.
  $$
If $x(\G1)$ and $x(\G2)$ have an interval in common, $\G{12}$ may even
disconnect in different pieces. The structure of these pieces
is easy determined
from a picture of the real part of the curve. Namely, on a square
representing $\G 1\times\G2$, the divisor $\{(P_1,P_2)\in\G i\times \G
j\mid x(P_1)=x(P_2)\}$ is drawn by counting points on the vertical lines
$x=$ constant, the only care one needs to take is that if $\G1$ (or
$\G2$) is closed, then an origin should be marked on it, and if one
passes this origin, one needs to pass over the corresponding edge of the
rectangle.
\smallskip
In the same way $\Gamma_{kk}$ is investigated by drawing the divisor
  $$ \left\{(P_1,P_2)\in\G i\times\G i\mid x(P_1)=x(P_2)\hbox{ and }\left(
    \eqalign{&P_1\neq P_2\hbox{ or,}\cr &P_1=P_2 \hbox{ is a ramification
    point of $x$}}\right)\right\}.
  $$
on a rectangle representing $\G i\times\G i$. Either triangle
cut off from the rectangle by its main diagonal then represents ${\G
i\times\G i\over S_2}$ and $\G{ii}$ is the complement of the divisor in
the triangle. For every $\G i$
such a piece is found and will be glued to $\G{00}$
precisely along the part of its boundary which comes from the diagonal
in the rectangle.
\medskip
In order to explain how $\G{00}$ is described, we recall the classical
picture of a (smooth, complete) algebraic curve $\GB$. An equation
$F(x,y)=0$ of such a curve defines an $m\colon1$ ramified covering map
to $\CP1$ by $(x,y)\mapsto x$, when $m$ is the degree of $F(x,y)$ in
$y$. This may be visualised by drawing concentric spheres (called {\it
sheets\/}), on which there are marked some non-intersecting intervals
(called {\it cuts\/}, every cut is equally present on all sheets).
The topology is such that if you are  walking on a sheet $i$ and pass a
cut $j$ (from a fixed side) then you move to a sheet $p_j(i)$, each
$p_j$ being a permutation of $\{1,\dots,m\}$. It is clear that the datum
of cuts and their corresponding permutations determines the topology of
the curve completely. Since each cut connects two ramification points
(of $x$), these cuts may, for a real curve, be taken on the real axis
and orthogonal to it.
\medskip
$\G{00}$ is now given as follows. Consider the described picture for the
smooth completion $\bar\GF$ of $\GF$. Clearly the conjugation map
interchanges the upper and lower hemispheres and is fixed on the
equator(s) $\{P\in\GBF\mid x(P)\in\Real\cup\infty\}$. It follows that
the open upper (lower) hemispheres give precisely $\G{00}$. A convenient way
to represent them is by drawing a disc for each upper hemisphere
and labelling the
different parts of the boundary which correspond to the horizontal and
vertical cuts. A moment's thought reveals that the different sheets are
to be connected along those lines which correspond to the vertical cuts,
while the pieces $\G{kk}$ are to be connected to the corresponding
horizontal cuts. This gives a topological model of
$\G{00}\cup\bigcup_{k=1}^s\G{kk}$ as a disc with holes. See the complete
version of this paper for an example (Pub.\ IRMA Lille 33 (1993));
it has been suppressed because it
contains several figures.
\subparg{Compactification of the complex invariant manifolds {\rm $\AFdC$}}
We now discuss the (smooth) compactification of the manifolds $\AFdC$.
There is one obvious and natural compactification, namely the compact
manifold $\Symd\GBF$, defined in a similar way as $\Symd\GF$; as above
$\bar\GF$ denotes the smooth compactification of $\GF$.
However $\AFdC$  has  the
disadvantage that none of the vector fields $X_{H_i}$ extends
holomorphically to it --- a compactification such that at least one of these
vector
fields extends in a holomorphic way to it, will simply be called {\it
good}. The interest in good compactifications is that it allows one to
integrate the corresponding vector fields in terms of theta functions,
or degenerations of theta functions, which are analytic, quasi-periodic
functions on $\C d$. The purpose of this paragraph is to show that even
for very simple choices of $F(x,y)$, a good compactification of $\AFdC$
does not
exist. We believe that this is true for almost all choices of $F(x,y)$.
A class of examples for which a good compactification {\it does} exist
is considered in the next section.
\smallskip
At first we compute, for fixed $F(x,y)$
how the vector fields $X_{H_i}$ behave on the compact
manifold $\Symd\GBF$, which relates to $\AFdC=\Symd\GF\setminus\DFd$ as
follows:
  $$ \Symd\GBF=\AFdC\cup\bar\DFd\cup\bar\EFd.
  $$
Here $\bar\DFd$ is the closure of $\DFd$ in $\Symd\GBF$ and
$\bar\EFd$ is a divisor whose irreducible components
$\bar\EFd(\infty_i)$ correspond to the points $\infty_i$ in
$\GBF\setminus\GF$, namely
  $$ \bar\EFd(\infty_k)=\left\{\langle \infty_k,P_2,\dots,P_d\rangle\mid
       P_k\in\GBF\hbox{ for } 2\leq k\leq d\right\}.
  $$
Each vector field $X_{H_i}$ being a
polynomial vector field on $\C{2d}$, it is holomorphic on $\AFdC$. We
determine its behaviour along the irreducible components of $\bar\DFd$
and $\bar\EFd$, which may be done by computing the order of vanishing of
$X_{H_i}$ at a generic point of each component, which in turn is done by
using local coordinates at such a point (see \referentie{GH}).
\Proposition{Every vector field $X_{H_i}$ has a simple pole along all
irreducible components of $\bar\DFd$. It has a zero of order $\rho_k$
along $\bar\EFd(\infty_k)$ (i.e., a pole of order $-\rho_k$ if
$\rho_k<0$), where
\smallskip\noindent
\centerline{\vbox{
  \halign{$\rho_k=\mu_k-\nu_k#,$ \hfil&\quad $\nu_k#0$&#\cr
  {+d+1}&<&,\cr
  {+1}&>&;\cr}}}
\smallskip\noindent
$\mu_k$ is the order (of vanishing) of ${\partial F\over\partial
y}(x,y)$ at $\infty_k$, and $\nu_k$ is the order of $x$ at $\infty_k$
(resp.\ the order of $x-x(\infty_k)$ if $x$ is finite at $\infty_k$.}
\propositionname{singul_vector_field}\Proof
We first write down the vector field $X_{H_i}$ at a
generic point $(u(\l),v(\l))\in\AFdC$; the genericity condition taken
here is that for
$\phi_{F,d}(u(\l),v(\l))=\langle(x_1,y_1),\dots,(x_d,y_d)\rangle$ all $x_i$ are
different and none of the points $(x_i,y_i)$ is a  ramification point of
$x$. Varying the point $(u(\l),v(\l))$ in a small neighborhood, each
$x_i$ gives a local coordinate on a neighborhood $U_i\subset\GF$ of
$(x_i,y_i)$ as well as a local coordinate on a neighborhood
$U\subset\AFdC$ of $\langle(x_1,y_1),\dots,(x_d,y_d)\rangle$.
Since on the one hand
the derivative of $u(\l)=\prod_{k=1}^d(\l-x_k)$ at $\l=x_j$ is
  $ X_{H_i} u(x_j)=-\prod_{l\neq j}(x_j-x_l)\,X_{H_i}
  x_j,
  $
while at the other hand, direct substitution in
(12)\nameformula{vector_field} gives
  $$ X_{H_i} u(x_j)={\partial F\over\partial
  y}(x_j,y_j)\sum_{k=0}^{i-1}u_kx_j^{i-k-1},
  $$
we find that
  $$X_{H_i}x_j=-\prod_{l\neq j}(x_j-x_i)^{-1}{\partial F\over\partial
      y}(x_j,y_j)\sigma_{i-1}(\hat x_j),
    \formula
  $$
\formulaname{cooflow}where
$\sigma_{i-1}(\hat x_j)$ is the $i-1$-th symmetric function in
$x_1,\dots,x_d$, evaluated at $x_j=0$.
\smallskip
The right hand side of
(20)\nameformula{cooflow} has at a generic point of $\bar\DFd$ a simple
pole, hence each vector field $X_{H_i}$ has a simple pole on (every
component of) $\bar\DFd$. The behaviour of $X_{H_i}$ along $\bar\EFd$ is
slightly more complicated since it depends on $F(x,y)$, and may even
behave differently on each components $\bar\EFd(\infty_k)$. For a
generic point in a neighborhood of a point of $\bar\EFd(\infty_k)$, let us
introduce coordinates $x_i$ as above. If we denote by $\mu_k$ and
$\nu_k$ the integers introduced in the Proposition, then clearly
$x_1$ is given on a neighborhood of
$\infty_k$ in terms of a
local parameter $t_1$ at $\infty_k$ as $x_1=t_1^{\nu_k}$ ($\nu_k<0$), or as
$x_1=c_1+t_1^{\nu_k}$ ($\nu_k>0$), depending on whether $x$ is infinite
in a neighborhood of $\infty_k$ or has a finite value $c_1\in\Complex$
at $\infty_k$; also ${\partial F\over\partial y}(x_1(t),y_1(t))
=t^{\mu_k}(f_1 +\otheta(t))$ with $f_1\neq0$.
We define for $2\leq j\leq
d$ local parameters $t_j$ (centered at $P_j$, which may be assumed to be
generic) by $x_j=x(P_j)+t_j$. Elementary substitution in
(20)\nameformula{cooflow} yields
  $$ \eqalign{X_{H_i}t_1&=t_1^{\rho_k}(c_i+\otheta(t_1)),\cr
              X_{H_i}t_j&=c_j+\otheta(t_1),\qquad (j=2,\dots,d);\cr}
  $$
where $\rho_k$ is defined as above. We conclude that
$X_{H_i}$ has a zero (resp.\ pole) of order
$\vert\rho_k\vert$ along $\bar\EFd(\infty_k)$ if $\rho_k\geq0$ (resp.
$\rho_k<0$).
\endproof
Thus we have shown that $\Symd\GBF$ is not a good compactification,
since all vector fields $X_{H_i}$ have at least a pole along $\bar\DFd$.
This divisor can be contracted in some cases, as we will show in the
next section. The following example shows that a good compactification
does not exist in general.
\Example
Let $F(x,y)=y^3+f(x)$, where the degree of $f$ is at least three, and
let $d=2$. To show that $\A^\Complex_{F,2}$ has no good compactification we use
some results about algebraic surfaces which can be found in
\referentie{Ha}. Suppose that $\bar{\cal A}$
is a good compactification of ${\cal A}^\Complex_{F,2}$ then $\bar{\cal
A}$ and $\Sym2\GBF$ are
birational; for surfaces this means that there exists a finite series
of monoidal transformations (also known as blow-up's) which transforms
$\bar{\cal A}$ into $\Sym2\GBF$. Then there exist (Zariski) open subsets
${\cal U}\subset\bar{\cal \A}$ and ${\cal V}\subset\Sym2\GBF$ to
which all these monoidal transformations restrict as isomorphisms and
the vector fields on $\cal U$ and $\cal V$ correspond exactly under this
isomorphism. In particular $\bar{\cal D}_{F,2}$ is entirely contained in the
complement of $\cal V$ and must be contracted by one of the monoidal
transformations, so at least we know that the genus of $\bar{\cal D}_{F,2}$
must be 0 (only $\CP1$'s can be contracted).

We may however compute the genus of $\bar{\cal D}_{F,2}$ directly.
Recall that it consists of the points $\langle P_1,P_2\rangle$ with
$x(P_1)=x(P_2)$ for which $P_1\neq P_2$ or $P_1=P_2$ is a ramification
point, so its smoothness is easy checked. However the map $x$
expresses $\bar{\cal D}_{F,2}$ as a $3\colon1$ cover of $\CP1$, ramified
at the $n=\deg f$ points $(x_i,y_i)$ for which $f(x_i)=0$ (and at
infinity if $n$ is not divisible by 3). So it has the same ramification
divisor as $\bar\GF$, hence $\hbox{genus}(\bar{\cal
D}_{F,2})=\hbox{genus}(\GBF)>0$, a contradiction.
\medskip
\Amplification
Summing up
(20)\nameformula{cooflow} over all $j$ (and for any $\varphi$)
we find that for any fixed integer $r<d$,
  $$\sum_{j=1}^d x_j^r{X_{H_i}^\varphi x_j\over\varphi(x_j,y_j)
            {\partial F\over\partial y}(x_j,y_j)}
              =-\sum_{k=0}^{j-1}u_k\sum_{j=1}^d
              x_j^{r+i-k-1}\prod_{l\neq j}{1\over x_j-x_l}
              =-\delta_{i+r,d}.
  $$
Therefore the $d$ variables
  $$ \chi_r=\sum_{i=1}^d{x_j^r X_{H_i}^\varphi x_j\over\varphi(x_j,y_j)
  {\partial F\over\partial y}(x_j,y_j)},\qquad r=0,\dots,d-1,
    \formula
  $$
\formulaname{linearisation}have
linear dynamics in time and lead to the explicit integration of the
vector fields $X_{H_i}$ along the real manifolds $\AFd$.
\medskip
\Amplification
In the one-dimensional case ($d=1$) the invariant
manifolds are punctured Riemann surfaces and have a unique
compactification. If the genus of such a Riemann surface
exceeds one, then it supports no holomorphic vector fields,
so for $d=1$ good
compactifications of $\AFdC$ rarely exist.
\endparg
\parg{The hyperelliptic case}
\sectionname{hyperelliptic}If
$F(x,y)$ is of the form $F(x,y)=y^2+f(x)$, for some
polynomial $f(x)$, then some simplifications occur; we call this the
{\it hyperelliptic} case, because $\GF$ is then a hyperelliptic curve.
We derive Lax equations for this case and discuss a hierarchy of
potentials on the plane,
the so-called H\'enon-Heiles hierarchy, which is intimely
related to the $d=2$ hyperelliptic case.
\subparg{Lax equations}
If $F(x,y)=y^2+f(x)$, then
  $$ {\partial F\over\partial y}({\l},v({\l}))=2v({\l}),
     \formula
  $$
\formulaname{hyperelliptic_F_derivative}and equations
(12)\nameformula{vector_field} can be written as Lax equations, i.e., they
can be written as a commutator in some Lie algebra (see e.g.
\referentie{Gr}), as given by the
following theorem.
\Theorem{The differential equations describing the vector fields for the
hyperelliptic case are written in the Lax form (with spectral parameter ${\l}$)
  $$X^\varphi_{H_i}A({\l})=\left\lbrack A({\l}),\plus{B_i({\l})}\right\rbrack,
  $$
where
  $$ A({\l})=\pmatrix{v({\l})& u({\l})\cr w({\l})&-v({\l})},
           B_i({\l})={\varphi(\l,v(\l))
                \over u({\l})}\plus{u({\l})\over
                {\l}^{d-i+1}}A(\l)\hbox{ and }
            w({\l})=-\plus{F({\l},v({\l}))\over u({\l})}.
  $$
\formulaname{hyperelliptic_Lax}The
spectral curve $\det(A(\l)-\m\;\Id)=0$, preserved by the flow of the
vector fields $X^\varphi_{H_i}$, is given by
$ \m^2+f(\l)=H_{F,d}(u(\l),v(\l))$.}
\Proof
If we define the polynomial $w({\l})$ as stated above, then equations
(12)\nameformula{vector_field} are easy rewritten as
  $$ \eqalign{X^\varphi_{H_i}u({\l})&=2\varphi(\l,v(\l))v(\l)
                \plus{u({\l})\over {\l}^{d-i+1}}
                -2u({\l})\plus{\varphi(\l,v(\l)){v({\l})\over
                u(\l)}\plus{u(\l)\over {\l}^{d-i+1}}},\cr
              X^\varphi_{H_i}v({\l})&=-\varphi(\l,v(\l))w({\l})\plus{u({\l})
                \over {\l}^{d-i+1}}+u({\l})\plus{\varphi(\l,v(\l)) {w({\l})
                \over u({\l})}\plus{u({\l})\over {\l}^{d-i+1}}}.}
    \formula
 $$
\formulaname{vector_field_hyperelliptic}upon using
(22)\nameformula{hyperelliptic_F_derivative}.
{}From
(23)\nameformula{vector_field_hyperelliptic} let us also compute
$X^\varphi_{H_i}w({\l})$ and remark
that the explicit dependence on $F$ disappears completely!
  $$ \eqalign{X^\varphi_{H_i} w(\l)
           &=-X^\varphi_{H_i}\plus{F(\l,v(\l))\over u(\l)},\cr
                &=-2\plus{{v(\l)\over u(\l)}X^\varphi_{H_i}v(\l)}+
                \plus{{F(\l,v(\l))\over u(\l)}{X^\varphi_{H_i}
                u(\l)\over u(\l)}},\cr
                &=2\plus{v({\l})\min{{w({\l})\varphi(\l,v(\l))
                \over u({\l})}\plus{u({\l})\over
                {\l}^{d-i+1}}}-
                w({\l}){X^\varphi_{H_i}u({\l})\over u({\l})}},\cr
                &=-2v(\l)\plus{\varphi(\l,v(\l)){w(\l)\over u(\l)}
                \plus{u(\l)\over {\l}^{d-i+1}}}
                +2w(\l)\plus{\varphi(\l,v(\l)){v(\l)\over u(\l)}
                \plus{u(\l)\over\l^{d-i+1}}}.}
  $$
This leads at once to the above Lax equations. The associated
spectral curve is computed as follows:
  $$ \eqalign{\det(A(\l)-\m\;\Id)&=\m^2-v^2(\l)+u(\l)\plus{f(\l)+v^2(\l)
                                     \over u(\l)},\cr
        &=\mu^2+f(\l)-u(\l)\min{f(\l)+v^2(\l)\over u(\l)},\cr
        &=\mu^2+f(\l)-H_{F,d}(u(\l),v(\l)).}
  $$
\endproof
For example, if we restrict ourselves to $d=1$ (i.e., one degree of freedom),
then $u({\l})={\l}+u_1,\,v({\l})=v_1$ and
  $$\eqalign{H_{F,1}(u_1,v_1)&=\left(v^2({\l})+f({\l})\right)\mod u({\l}),\cr
        &=\left(v_1^2+f({\l})\right)\mod({\l}+u_1),\cr
        &=v_1^2+f(-u_1),\cr}
  $$
and $\{\cdot,\,\cdot\}_1^1$ is the standard bracket on $\R2$,
so we find that for $\varphi=1$
the hyperelliptic case in one degree of freedom corresponds
exactly to the case of polynomial potentials on the line.
\medskip
\Amplification
In the special case that $\varphi=1$ and $d=\hbox{genus}(\GF)$ one
obtains the so-called
odd or even master systems, according to whether the degree
of $f(x)$ is odd or even. The odd master system was introduced by
Mumford in \referentie{Mu} and his construction was adapted by us to
obtain the even master system (see \referentie{V}). Both systems are
known to be algebraic completely integrable. Indeed, the Abel map
(recalled in the next paragraph)
linearises the vector fields on the Jacobian of the spectral curve
$\det(A(\l)-\m\;\Id)=0$, as follows from
(21)\nameformula{linearisation}. In this special case we may rewrite the matrix
$\plus{B_i(\l)}$ as
  $$ \plus{B_i(\l)}=\plus{A(\l)\over\l^{d-i+1}}+\pmatrix{0&0\cr b_i&0},
  $$
where $b_i=-u_i$ or $b_i=-u_i\l+2u_1u_i-u_{i+1}$, according to
whether the degree of $f(x)$ is odd or even\footnote{2}{We
killed in the latter case the coefficient of $x^{g+1}$ in $f(x)$,
precisely as we did in \referentie{V}.}, showing that these systems coincide
indeed with the
master systems in \referentie{V}.
\subparg{$\AFdC$ as strata of hyperelliptic Jacobians}
Recall from \referentie{GH} that a complex (algebraic) torus is
associated to any (complex, smooth, complete) algebraic curve
$\GB$ as follows: choose a base $\{\omega_1,\dots,\omega_g\}$ for the
holomorphic differentials and a symplectic base
$\{A_1,\dots,A_g,B_1,\dots,B_g\}$ for $H_1(\GB,\Z)$, symplectic meaning
here that $A_i\cdot A_j=B_i\cdot B_j=0$ and $A_i\cdot B_j=\delta_{ij}$.
If we denote  $\vec\omega=(\omega_1,\dots,\omega_d)$ then the lattice
  $$ \Lambda_{\GB}=\hbox{\rm span }
       \left\{\int_{A_i}\vec\omega\,, \int_{B_i}\vec\omega\mid 1\leq
  i\leq d\right\}
  $$
is of rank $2d$ and a complex torus (which can be shown to be
independent of the choices made) is defined by $\Jac(\GB)=\C
g/\Lambda_{\GB}$, the so-called {\it Jacobian} or $\GB$.
Fixing any base point $P_0\in\GB$
there is for each $d\in\N$ a well-defined holomorphic
map $A_d\colon\Symd\GB\to\Jac(\GB)$ defined
by
  $$ A_d(\langle P_1,\dots,P_d\rangle)=\sum_{i=1}^d\int_{P_0}^{P_i}
  \vec\omega\mod\Lambda_{\GB},
  $$
classically known as {\it Abel's map\/}; Abel's Theorem says that $A_d(\langle
P_1,\dots,P_d\rangle)=A_d(\langle Q_1,\dots,Q_d\rangle)$ if and only if
there exists a meromorphic function on $\GB$ with zeros at
the points $P_1,\dots,P_d$ and poles at $Q_1,\dots,Q_d$.
\smallskip
The complex invariant manifolds $\AFdC$ behave well with respect to this
map in the hyperelliptic case, as is shown in the following proposition.
For a smooth curve $\Gamma\subset\C2$ we will denote its completion
(i.e., smooth compactification) by $\GB$. In the case of hyperelliptic
curves $y^2+f(x)=0$ this completion $\GB$ is obtained by adding to
$\Gamma$ one or two points, depending on whether the degree of $f(x)$ is
odd or even; these points will be denoted by $\infty$, resp.\ $\infty_1$
and $\infty_2$. Remark that if $F(x,y)=y^2+f(x)$ then $\GF$ is smooth
(or, equivalently, $\AFdC$ is smooth) if
and only if $f(x)$ has no multiple roots.
\Proposition{In the hyperelliptic case $F(x,y)=y^2+f(x)$,
the complex invariant manifold $\AFdC$ is for $d\leq g$ biholomorphic to a
(smooth) affine part of a distinguished $d$-dimensional subvariety $W_d$
of $\Jac(\GBF)$, namely
\smallskip
\noindent
\centerline{\vbox{
  \halign{$\AFdC\,\cong\,#\hfil\qquad$&$\deg f(x)$ #,\hfil\cr
        W_d\setminus W_{d-1}&odd\cr
        \noalign{\smallskip}
        W_d\setminus \left(W_{d-1}\cup(\vec e+W_{d-1})\right)&even\cr}}}
\smallskip\noindent
where $\vec e\in\Jac(\GBF)$ is given by $\vec
e=A_1(\infty_1)-A_1(\infty_2)=\int_{\infty_2}^{\infty_1}
\vec\omega\mod\Lambda_{\GBF}$.
Also
  $$ \eqalign{W_g&=\Jac(\GBF),\cr
              W_{g-1}&=\hbox{ theta divisor } \Theta\subset\Jac(\GBF),\cr
                      &\qquad\vdots\cr
              W_1&=\hbox{ curve } \GBF \hbox{ embedded in } \Jac(\GBF),\cr
              W_0&=\hbox{ origin of } \Jac(\GBF).}
  $$
}
\Proof
We prove the proposition only for the case in which $\deg f(x)$ is odd.
We choose $\infty$ as the base point for the Abel map and define $W_k$
for $k=1,\dots,g$ as $W_k=A_k(\Sym k\GBF)$. By a theorem due to Jacobi
$W_g=\Jac(\GBF)$ and by Riemann's Theorem, $W_{g-1}$ is (a translate of)
the Riemann theta divisor (see \referentie{GH}). Clearly for each $k\leq
g,\, W_{k-1}$ is a divisor in $W_k$ and, by another theorem of Riemann,
$W_k\setminus W_{k-1}$ is smooth. We claim that
  $$ A_d(\Symd\GF\setminus\DFd)=W_d\setminus W_{d-1},
  $$
more precisely $A_d$ realises a holomorphic
bijection between these smooth varieties. Namely,
  $$ \eqalign{\langle P_1,\dots,P_d\rangle&\in\Symd\GF\setminus\DFd\cr
           &\hbox{ iff }\forall i\;P_i\neq\infty\hbox{ and } \exists
           i\neq j\,\colon x(P_i)=x(P_j)\Rightarrow\left(
             \eqalign{&P_i=P_j\hbox{ and}\cr &P_i\hbox{ is not a
             ramification point of $x$}}\right)\cr
          &\hbox{ iff } A_d(P_1,\dots,P_d)\notin W_d\setminus W_{d-1},}
  $$
where we used Abel's Theorem in the last step. It follows that
$\Symd\GF\setminus\DFd$ and $W_d\setminus W_{d-1}$ are biholomorphic, hence
by Proposition
7\nameproposition{biholomorphism}, $A_d\circ\phi_{F,d}$ is a
biholomorphism and the manifolds
$\AFdC$ and $W_d\setminus W_{d-1}$ are biholomorphic.
\endproof
Using the results (and notation) of Sect.\ 3.4 we can determine very
precisely how the vector fields $X_{H_i}$ behave on $\Symd\GBF$. If
$\deg f(x)$ is odd, then $\nu(\infty)=-2$ and $\mu(\infty)=-2g-1$, so
that $\rho(\infty)=2(d-g)$. In the even case, we have that for $i=1,2$,
$\nu(\infty_i)=-1,\,\mu(\infty)=-g-1$ and $\rho(\infty)=d-g$.
Recall that these vector fields have in both  cases a
simple pole along $\bar\DFd$. If $d=g$, Abel's Theorem implies
that the Abel map contracts $\bar\DFd$ into something lower
dimensional; this fact, combined with the preceeding computation and
Proposition
9\nameproposition{singul_vector_field}, yields a
holomorphic vector field on the complex torus $A_g(\Symd\GBF)$ (it can
have no poles on $A_g(\bar{\cal D}_{F,g)}$  since this is of codimension two).
This explains why the master systems are algebraic completely integrable
(see Amplification 6). For $d>g$ we identify two $d$-tuples in $\Symd\GBF$
when both contain a pair of points of the form $(x_1,y_1),\,(x_1,-y_1)$
and have their other $d-2$ points equal. When a smooth manifold is
obtained, the vector fields $X_{H_i}$ are again holomorphic on them and
they are integrated in terms of degenerations of theta functions.
\subparg{The H\'enon-Heiles hierarchy}
It was found by Ramani (see \referentie{DGR}) that the integrable
H\'enon-Heiles potential $V_3=8q_2^3+4q_1^2q_2$ is part of a hierarchy of
integrable potentials
  $$ V_n=\sum_{k=0}^{[n/2]}2^{n-2k}{n-k\choose k}q_1^{2k}q_2^{n-2k}.
  $$
Namely, the energy $E_n=\left(p_1^2+p_2^2\right)/2+V_n$ has an extra
invariant given by
  $$ G_n=-q_2p_1^2+q_1p_1p_2+q_1^2V_{n-1},
  $$
as is checked immediately by direct computation. These potentials have
moreover the special property that they can be superimposed freely in
the sense that any linear combination of them gives an integrable
potential. The case $n=3$ was studied in \referentie{AvM4} and the case
$n=4$ in \referentie{V} (it was called the quartic potential there).
In fact, in \referentie{V} we constructed a map which relates this
quartic potential to the two-dimensional even master system. This map
will prove useful to understand the geometry of the whole H\'enon-Heiles
hierarchy. Namely define a map
$T\colon\C4\to\C4$ by
  $$T(q_1,q_2,p_1,p_2)=(\l^2-2q_2\l-q_1^2,\,-2p_2\l-2q_1p_1),
  $$
which is invariant for the action of $\Z_2$ on each complex invariant manifold
  $$ \A_{eg,n}^{\Complex}=\left\{P\in\C4\mid
           E_n(P)=e,\,G_n(P)=g\right\},
  $$
the action being given by $(q_1,q_2,p_1p_2)\mapsto(-q_1,q_2,-p_1,p_2)$.
It is fixed point free on $\A_{eg,n}^\Complex$ if $g\neq0$.
\Proposition{The map $T\colon\C4\to\C4$ given by
  $$T(q_1,q_2,p_1,p_2)=(\l^2-2q_2\l-q_1^2,\,-2p_2\l-2q_1p_1),
  $$
restricts to an unramified $2\colon1$ covering map on each invariant
manifold $\A_{eg,n}^\Complex$ (with $g\neq0$) and this restriction is
onto $\A_{F,2}^\Complex$, where $F$ is given by
  $$ F(x,y)=y^2+8\left(x^{n+2}-ex^2-gx\right),
  $$
and $\GF$ has genus $\left[{n+1\over2}\right]$. Therefore, if $n$ is odd
(resp.\ even)
then $\A_{eg,n}^\Complex$ is an unramified cover of the complement of
one (resp.\ two) curves, isomorphic to $\GBF$, in the $W_2$ stratum of
$\Jac(\GBF)$. The restriction $\tilde T$ of $T$ to $\A_{eg,n}^\Complex$
maps also the vector fields $X_{E_n}$ and $X_{G_n}$ to
(a multiple of) $X_{H_1}$ and $X_{H_2}$, and leads to the Lax
equations
$X_{E_n}A(\l)={1\over2}[A(\l),B_n(\l)]$,
for the H\'enon-Heiles hierarchy, where
  $$ A(\l)=\pmatrix{-2p_2\l-2q_1p_1 &\l^2-2q_2\l-q_1^2\cr
                    4p_1^2-\sum_{i=0}^nV_i\l^{n-i} &2p_2\l+2q_1p_1},
    \quad
    B(\l)=\pmatrix{0&1\cr-8\sum_{i=1}^{n-1}V'_j\l^{n-i-1}&0}
  $$
and  $V_j'={\partial V_j\over\partial q_2}(q_1,q_2)$.}
\propositionname{GeoHHH}\Proof
Let us fix values $e,\, g$ and denote by $\tilde T$ the restriction of
$T$ to $\A_{eg,n}^\Complex$. We show that $\tilde T$ maps
$\A_{eg,n}^\Complex$ in $\A_{F,2}^\Complex$, when $F(x,y)$ is defined as
$F(x,y)=y^2+8(x^{n+2}-e_nx^2-g_nx)$. To show this, let
$(q_1,q_2,p_1,p_2)\in\A_{eg,n}^\Complex$ and
let $u(\l)=\l^2-2q_2\l-q_1^2$ and $v(\l)=-2p_2\l-2q_1p_1$.
Then the equality
  $$ {F(\l,v(\l))\over u(\l)} = 8\sum_{i=0}^nV_i\l^{n-i}-4p_1^2,
     \formula
  $$
\formulaname{quotient}follows immediately from
  $$ \eqalign{\left(\sum_{i=0}^{n-1}V_i\l^{n-i}\right)(\l^2-2q_2\l-q_1^2)
          &=\sum_{i=-2}^{n-3}V_{i+2}\l^{n-i}-2q_2\sum_{i=-1}^
          {n-2}V_{i+1}\l^{n-i}-q_1^2\sum_{i=-1}^{n-1}V_i\l^{n-i},\cr
          &=\l^{n+2}+\sum_{i=-1}^{n-2}(V_{i+2}-2q_2V_{i+1}-q_1^2V_i)\l^{n-i}
          -V_n\l^2-q_1^2V_{n-1}\l,\cr
          &=\l^{n+2}-V_n\l^2-q_1^2V_{n-1}\l,}
  $$
where we used in the last step the recursion formula
  $$ V_{i+2}=2q_2V_{i+1}+q_1^2V_i
     \formula
  $$
\formulaname{recursion}for
the potentials $V_i$ (valid for $i\geq-1;\,V_{-1}=0$). It follows
that $\tilde T$ maps $\A_{eg,n}^\Complex$ indeed in $\A_{F,2}^\Complex$.
Clearly $\tilde T$ is onto and unramified.
\medskip
To obtain a Lax pair, let $\varphi(x,y)=1$ and compute  the entries in
$\plus{B_i(\l)}$ as given by
(22)\nameformula{hyperelliptic_Lax}. The only non-trivial element in
$B_i(\l)$ is $\plus{w(\l)\over u(\l)}$, where
$w(\l)=4p_1^2-8\sum_{i=0}^nV_i\l^{n-i}$, as follows from the definition
of $w(\l)$ in
(22)\nameformula{hyperelliptic_Lax} and
(24)\nameformula{quotient}. As in the preceeding calculation we get
  $$(\l^2-2q_2\l-q_1^2)\sum_{j=1}^{n-1}V_j'\l^{n-j-1}
     =2u(\l)\sum_{i=0}^{n-1}\sum_{i=0}^{n-1}V_i\l^{n-i}+\hbox{ (polynomial
     of degree $\leq1$)},
  $$
by using the relation
  $$ V_{i+2}-2q_2V'_{i+1}-q_1^2V'_{i}=2V_{i+1},
  $$
which is the derivative of
(25)\nameformula{recursion} with respect to $q_2$. From this representation
it is seen at once that $\tilde T_*X_{E_n}={1\over2}X_{H_1}$. Similarly one
shows that $T_*X_{G_n}$ is a multiple of $X_{H_2}$.
\endproof
Using the results of Paragraph 3.3, the topology of the real invariant
manifolds as well as the bifurcations of the H\'enon-Heiles hierarchy
can be determined, in analogy
with \referentie{Ga}, where this is done for the case $n=3$ (the
H\'enon-Heiles potential).
\medskip
\Amplification
As we learned from V. Kuznetsov, the H\'enon-Heiles hierarchy has a
higher dimensional generalisation, which consists of a family of
potentials on $\R d$, defined by a recursion relation which generalises
(25)\nameformula{recursion}, namely let $B$ and $A_1,\dots,A_{d-1}$ be
arbitrary parameters, the $A_i$ being all different. Then the potentials
are defined by
  $$ V^{(d)}_{i+2}=2(q_d-B)V^{(d)}_{i+1}+\sum_{k=1}^{d-1}\sum_{j=0}^i(-1)^j
  q_k^2V^{(d)}_{i-j}A^j_k;
  $$
the H\'enon-Heiles hierarchy discussed above corresponds then to the
case $d=2,\,A_1=B=0$. Using the results obtained in \referentie{EEKL},
it is easy to construct the generalisation of our map $T$ and to
generalise Proposition
12\nameproposition{GeoHHH}, i.e., to prove
that for the $n$-th member $V^d_n$ of the hierarchy $(n\geq3)$,
the complex invariant manifolds are $2^{d-1}\colon1$ unramified covers
of (an affine part of) the $W_d$ stratum of the hyperelliptic Jacobian
$\Jac(\GBF)$, where
  $$F(x,y)=y^2+\pi(x)\left(16x^{n-2}(x+B)^2+8e_n+
    \sum_{i=1}^{d-1}f_i{\pi(x)\over
    x+A_i}\right),\qquad\pi(x)=\prod_{i=1}^{d-1}(x+A_i),
  $$
which defines a hyperelliptic curve of genus $\lbrack{n-3\over2}\rbrack+d$. It
leads
also in a natural way to Lax equations for this hierarchy.
\ackn{I wish to thank P. Bueken, L. Haine, V. Kutznetsov and P. van
Moerbeke for valuable and stimulating discussions which
contributed to this paper. The support of the Max-Planck-Institut f\"ur
Mathematik is also greatly acknowledged.}
\endparg
\referenties
\refbook{A}{Arnold, V.}
           {Mathematical Methods of Classical Mechanics}
           {Springer-Verlag 1978}
\refbook{AM}{Abraham, R., Marsden, J.}
           {Foundations of Mechanics}
           {Benjamin/Cummings Publishing Company 1978}
\refbook{AN}{Arnold, V., Novikov, S.P. (eds)}
           {Dynamical Systems IV}
           {Springer-Verlag 1990}
\refbook{AvM1}{Adler, M., van Moerbeke, P.}
               {Algebraic Completely Integrable Systems: a systematic approach}
               {Perspectives in Mathematics, Academic Press (to appear
               in 1994)}
\refart{AvM2} {Adler, M., van Moerbeke, P.}
               {Completely Integrable Systems, Euclidean Lie Algebras, and
Curves}
               {Adv.\ Math.}
               {38}
               {267--317 (1980)}
\refart{AvM3} {Adler, M., van Moerbeke, P.}
               {Linearisation of Hamiltonian Systems, Jacobi Varieties and
Representation Theory}
               {Adv.\ Math.}
               {38}
               {318-379 (1980)}
\refart{AvM4} {Adler, M., van Moerbeke, P.}
     {The Kowalewski and H\'enon-Heiles Motions as Manakov Geodesic
     Flows on $SO(4)$ --- a Two-Dimensional Family of Lax Pairs}
     {Comm.\ Math.\ Phys.}
     {113}
     {659--700 (1988)}
\refart{DGR}{Dorizzi, B., Grammaticos, B., Ramani, A.}
     {Painlev\'e Conjecture Revisited}
     {Phys.\ Rev.\ Lett.}
     {49}
     {1539--1541 (1982)}
\refpre{EEKT}
       {Eilbeck, J.C., Enol'skii, V.Z., Kuznetsov, V.B., Tsiganov, A.V.}
       {Linear R-matrix algebra for classical separable systems}
\refart{Ga}{Gavrilov, L.}
     {Bifurcations of invariant manifolds in the generalized
     H\'enon-Heiles system}
     {Phys.\ D}
     {34}
     {223--239 (1989)}

\refart{Gr} {Griffiths, Ph.}
           {Linearising Flows and a Cohomological Interpretation of Lax
Equations}
           {Amer.\ J. Math.}
           {107}
           {1445--1483 (1985)}
\refbook{GH}{Griffiths, Ph., Harris, J.}
           {Principles of Algebraic Geometry}
           {Wiley-Interscience, New-York 1978}
\refbook{GS}{Guillemin, V., Sternberg, S.}
           {Symplectic techniques in physics}
           {Cambridge Univ.\ Press 1984}
\refbook{Gu}   {Gunning, R.}
               {Lectures on Riemann surfaces, Jacobi varieties}
               {Princeton Mathematical notes 1976}
\refbook{H} {Hartshorne, R.}
            {Algebraic Geometry}
            {Springer-Verlag 1977}
\refart{K}  {Krichever, I.}
          {Methods of algebraic geometry in the theory of non-linear equations}
               {Russian Math.\  Surveys}
               {32}
               {185--213 (1977)}
\refbook{LB}{Lange, H., Birkenhake, Ch.}
               {Complex Abelian Varieties}
               {Springer-Verlag 1992}
\refbook{M}   {Mumford, D.}
               {Tata Lectures on Theta 2}
               {Birkh\"auser 1984}
\refbook{Mo}{Moser, J.}
           {Various Aspects of Integrable Hamiltonian Systems}
           {Progress in Mathematics 8, Birkha\"user 1980}
\refart{P}{Painlev\'e, P.}
          {Sur les fonctions qui admettent un th\'eor\`eme d'addition}
          {Acta Math.}
          {25}
          {1--54 (1902)}
\refbook{S}{Silhol, R.}
      {Real Algebraic Surfaces}
      {Springer-Verlag 1989}

\refart{V} {Vanhaecke, P.}
               {Linearising two-dimensional integrable systems and the
               construction of action-angle variables}
               {Math.\ Z.}
               {211}
               {265--313 (1992)}
\refart{VN}    {Veselov, A., Novikov, S.}
               {Poisson brackets and complex tori}
               {Proc.\ Steklov Inst.\ Math.}
               {3}
               {53--65 (1985)}

\end